\documentclass[12pt,a4wide]{article}
\usepackage{graphicx}     
\usepackage{amssymb}  
\begin{document}      
     
{\Large \bf EM wave propagation in two-dimensional photonic crystals: 
a study of anomalous refractive effects}\\
\vspace{0.5cm}

{\large S. Foteinopoulou$^{1}$ and C. M. Soukoulis$^{1,2}$}\\
\vspace{0.5cm}
{\it $^{1}$ Ames Laboratory-USDOE and Dept. of Physics and Astronomy, Iowa State University, Ames IA, 50011,USA}\\
{\it $^{2}$ Research Center of Crete, Heraklion, Crete 71110, Greece}\\
\vspace{1cm}

{We systematically study a collection of refractive phenomena that can possibly occur at the interface of a two-dimensional photonic crystal, with the use of the wave vector diagram formalism. Cases with a single propagating beam (in the positive or the negative direction) as well as cases with birefringence were observed. We examine carefully the conditions to obtain a single propagating beam inside the photonic crystal lattice. Our results indicate, that the presence of multiple reflected beams in the medium of incidence is neither a prerequisite nor does it imply multiple refracted beams. We characterize our results in respect to the origin of the propagating beam and the nature of propagation (left-handed or not). We identified four distinct cases that lead to a negatively refracted beam. Under these findings, the definition of phase velocity in a periodic medium is revisited and its physical interpretation discussed. To determine the ``rightness'' of propagation, we propose a wedge-type experiment. We discuss the intricate details for an appropriate wedge design for different types of cases in triangular and square structures. We extend our theoretical analysis, and examine our conclusions as one moves from the limit of photonic crystals with high index contrast between the constituent dielectrics to photonic crystals with low modulation of the refractive index. Finally, we examine the ``rightness'' of propagation in the one-dimensional multilayer medium, and obtain conditions that are different from those of two-dimensional systems.\\
PACS~numbers:~78.20.Ci, 41.20.Jb,42.25.-p, 42.30.-d \\}

\vspace{1.5cm}   
{\bf I. INTRODUCTION}
\vspace{0.2cm}  
\par
Photonic crystals (PC's) are dielectric structures with two- or three- dimensional periodicity. They are known as the semiconductor counterpart for light, since they exhibit the ability,  ---when engineered appropriately--- to mold and control the propagation of EM waves. Among their unusual properties lies their ability to exhibit a wide variety of anomalous refractive effects, which recently attracted a great deal of interest, both theoretically \cite{Notomi,Gralak,Joann,Joann2,fotein} and experimentally \cite{kosaka,ertugul}. The observed refractive effects can be quite complicated, and, in most cases, the direction of the propagating signal cannot be interpreted with the use of a simple Snell-like formula. In particular, Kosaka et al. \cite{kosaka} observed a large swung of angle for the refracted beam, for a small angle of incidence. They called this effects the ``superprism phenomenon.'' 
\par  
Anomalous refractive phenomena are known in the field of 
optics and are commonly associated with anisotropy in the optical 
properties of the material (permittivity)\cite{born}. Two propagating solutions exist, having a different dispersion relation. One of them is extraordinary, i.e., non-spherical. As a result, in some cases two refracted beams are observed in these media, a phenomenon known as ``birefringence''\cite{born,mathieu}. Numerous studies \cite{Russell1,Russell2,Russell3,Russell4,Zengerle} on diffraction gratings, essentially the one-dimensional (1D) counterpart for the PC structures, led to the observation of a vast variety of anomalous refracted effects, including  ``birefringence.'' These systems have undergone extensive and systematic study \cite{Russell1,Russell2,Russell3,Russell4,Zengerle}, based on the wave vector diagram formalism. This formalism was proven to be an excellent tool in explaining the unusual refractive properties for the 1D diffraction grating system. The reader can find a didactic description of these diagrams and their use in Refs. \cite{Russell1} and \cite{Russell2}. 
\par  
Despite the recent  interest focused on the superrefractive effects in two-dimensional crystals, a systematic study is certainly lacking in the literature for these systems and only a few effects were studied and discussed \cite{Notomi,Gralak,Joann,fotein,kosaka,kosaka2}. So far, all studies on the two-dimensional (2D) photonic crystals are restricted to effects that can be explained with the study of the corresponding propagating modes within the first Brillouin zone (BZ). However, as we will demonstrate in this paper, a class of unusual propagation phenomena in PC's can be explained only by a careful study of all allowed propagation modes in all zones. Our analysis indicates that, contrary to one's intuition, the multiplicity of reflected beams in the incoming medium does not necessarily imply the presence of multiple beams in the PC medium and vice versa. In this context, we also investigate carefully the conditions necessary to obtain single beam propagation inside a 2D square or triangular PC lattice.
\par 
The anomalous refractive effects observed in both the 1D grating and PC literature include cases where the light bends ``the wrong way,'' i.e., is refracted negatively at the air-PC interface. Such a phenomenon was observed and widely discussed in the left-handed materials literature \cite{Veselago,Smith,Pendry,Soukoulis}. In the left-handed medium (LHM), homogeneous \cite{Veselago} or composite \cite{Smith},  the electric field vector {\bf E}, the magnetic field vector {\bf H}, and the wave vector {\bf k} form a left-handed set of vectors. The sign of the product $\bf S\cdot k$ {\hspace{1mm}} ---{\bf S} being the Poynting vector--- {\hspace{1mm}} reflects the sign of the ``rightness'' for the system \cite{Veselago}, and is negative for the left-handed medium (LHM). It is also customary to refer to a left-handed propagating wave as a backwards wave \cite{belov}. The refractive index for such a medium was calculated with the use of the scattering data and was found to be unambiguously negative \cite{Soukoulis}. However, the characterization of the left-handed or right-handed nature of propagation is a point somewhat overlooked both in the 1D gratings and PC literature. Only in Ref. \cite{fotein} the sign of the product $\bf S\cdot k$ \cite{note}, with {\bf k} in the first BZ, was determined for a two-dimensional hexagonal PC with a finite difference time domain simulation (FDTD) \cite{taflove}. The simulation experiment, performed on a wedged  PC structure, is in accordance with the UCSD experiment \cite{Smith2,Boeing}. In the latter, the negative index was experimentally verified for the traditional composite LHM. Left-handed behavior in photonic crystals relates to the origin and nature of a certain propagating beam. We intend to study this for different cases with the wave vector diagram formalism. In any case, the assignment of a proper refractive index, should carry the information regarding the rightness of the PC medium in its sign and be consistent with the left-handed literature. 
\par        
Moreover, the phase velocity for an EM wave propagating in a periodic structure is a subject of some controversy in the literature. Yariv defined  the phase velocity for a propagating EM wave in the 1D layered medium as the phase velocity that corresponds to the dominant plane wave component \cite{yariv}. Recently the phase velocity has been associated with the Bloch's crystal momentum {\bf k}, where {\bf k} is in the first BZ \cite{Notomi,fotein,kosaka3}. Specifically, in Ref. \cite{Notomi} the phase velocity and appropriate phase index were discussed for both limits of index contrast between constituent dielectrics (high and low). Considering this controversy, the subject of phase velocity in a periodic medium should be revisited. It is certainly worthwhile to reexamine the physical meaning of each definition in both limits of refractive index modulation (high and low).
\par  
In an attempt to make the study of the PC system simpler, in some cases the PC system was homogenized  appropriately with the use of an effective medium theory \cite{datta}. However these theories mainly apply to the long wavelength limit.  However, as we will show in our subsequent analysis, for some cases that lie in the higher bands, it is still possible to characterize the refractive and propagation properties with an effective index $n(\omega)$ under certain conditions. It is important to examine carefully such conditions, since the study of the PC can be greatly simplified. We will see that in these cases, both phase and energy velocities can be derived by simple formulas.
\par  
In this work, we attempt a systematic study for the anomalous refractive phenomena occurring in two-dimensional PC systems. We focus on various cases that have substantially different origins and nature. The characteristics of each case are analyzed. For this purpose, we discuss the properties of phase and energy velocities, as well as types of propagation (left- or right-handed). In particular, in Sec. II we present four distinct cases of anomalous refractive effects, where a negatively refracted beam is present. We explain and analyze the origin of the refracted beam with the wave vector diagrams in Sec. III. Using the same formalism,  we characterize birefringent phenomena in photonic crystals that we observed. We also discuss how these relate to the presence of Bragg reflections  in the medium of incidence. We discuss all relevant properties for an EM wave propagating in the crystal. In particular, we define appropriately a phase velocity, calculate it numerically, and discuss the meaning of the associated effective phase index in  Secs. IV , V and VI,  respectively. Moreover, in Sec. VI we discuss the conditions necessary to obtain single beam propagation. We derive expressions for both group and energy velocities, and show  their equality in Sec. VII. In Sec. VIII, we discuss the group refractive index associated with the group velocity. In Sec. IX we focus our discussion on the left- or right-handed nature of propagation. In this section, we discuss the details of the appropriate wedge experiment design, that unveils the sign of ``rightness'' for propagation inside the PC for different cases of triangular and square lattices. In Sec. X, we discuss the validity of our theoretical analysis as one moves from the limit of high-index modulated crystals to the limit of photonic crystals with low index modulation. Finally, we make a comparison between the two-dimensional PC medium and the 1D layered medium in Sec. XI. We present our conclusions in Sec. XII. 
\par
\vspace{1.5cm}
{\bf II. ANOMALOUS REFRACTIVE PHENOMENA AT THE AIR PC INTERFACE}
\vspace{0.2cm}
\par 
We present in Fig. 1 four characteristic cases where a negatively refracted beam appears when light is incident at  a PC slab interface. The cases shown in Fig. 1 basically outline the different possible reasons for which a negatively refracted beam can appear inside the photonic crystal.  In the case of Fig. 1(b), two distinct beams propagating in opposite directions (positive and negative) are present (birefringence). To study the various super-refractive effects, we employed the Finite Difference Time Domain technique (FDTD) \cite{taflove,yee} with Perfect Matched Layer (PML) \cite{Berenger} boundary conditions. We study various triangular PC structures of dielectric cylindrical pillars in air for the H (TE)-polarization case (magnetic field aligned along the cylinder's axis). Whenever possible, we used the value of 12.96 for the dielectric constant and r=0.35 for the radius of the rods for consistency and comparison with the results in Ref. \cite{fotein} and the results of Notomi \cite {Notomi}. However sometimes for the purpose of isolating and observing clearly specific effects, it becomes necessary to employ PC structures with different parameters. The presence of a negatively refracted beam is clear in all four cases as seen in Figs. 1(a)-(d). Before we expand our analysis, we discuss the wave vector diagrams. Careful use of such diagrams in the PC system can always explain/determine the direction(s) of the refracted beam(s). Then, we will be able to comment on the nature/origin of each different superrefractive effect shown in Fig. 1.   
\par  
\vspace{1.5cm}
{\bf III. WAVE VECTOR DIAGRAMS AND INTERPRETATION OF THE FDTD RESULTS}
\vspace{0.2cm}
\par
The wave vector diagram contains the equifrequency surfaces (EFS) for the photonic crystal that apply for the frequency of operation. Actually, for our two-dimensional system the surfaces reduce to contours. These contours consist of all allowed propagation modes in wave vector space that exist in the PC system for a certain frequency.  One or multiple contours can be relevant for a certain frequency, depending on the number of bands corresponding to the frequency of interest. To isolate the different effects, we focus our study on cases with only one band corresponding to the frequency of interest. Thus, we have only a single EFS within the first BZ in {\bf k}-space, closed or broken (with six-fold symmetry for the hexagonal lattice). The corresponding analogue of the EFS for the electronic case would be the Fermi surface. However, unlike electronic states that exist inside the crystal, an EM wave propagating inside the photonic crystal is excited with an EM wave incident from the outside. This implies the following: 1) the propagating state inside the crystal has the same frequency as the frequency of the source. 2) the causal direction of propagation inside the crystal points away from the source. 3) the wave vector of the propagating wave inside the crystal is subject to additional restrictions, imposed by the boundary. Specifically, the parallel component of the wave vector is given by the following formula \cite{sakoda}.\\  
\begin{eqnarray} 
k_{\|,m}=\frac{\omega}{c}\sin \theta_{inc}+\frac{2\pi m}{b_{str,cut}} \hspace{0.5mm},
\end{eqnarray}
where we consider EM waves incident from air with angle $\theta_{inc}$ measured from the surface normal. m is an integer equal to $0,\pm 1,\pm2$ and  
$b_{str,cut}$ represents the periodicity of the interface. For different interface cuts and lattice arrangements, we have the following cases:
\begin{displaymath}
 b_{str,cut}=\left \{ \begin{array}{ll} a & \textrm{for triangular cut along $\Gamma K$}\\ 
\sqrt 3 a & \textrm{for triangular cut along $\Gamma M$ \hspace{1mm},\hspace{5mm}  (A1)}\\
\sqrt 2 a & \textrm{for square cut along $\Gamma M$ }\\
a & \textrm{for square cut along $\Gamma X$}
\end{array}\right.
\end{displaymath}
with $a$ being the lattice constant.
\par  
Basically, formula (1) is the generalization of the phase matching condition at a periodic boundary \cite{oldano}. So, the $\bf k_{\|}$ conservation condition as expressed with Eq. (1) is an integral part of the wave vector diagram. Diagrammatically, Eq. (1) can be represented by m parallel lines, all perpendicular to the interface and separated by $2 \pi/b_{str,cut}$. We refer to these lines as construction lines in accordance with existing nomenclature in the literature (see for example Ref. \cite{Russell1}).  

The EM wave that propagates in the photonic crystal is a superposition of many plane waves, called Floquet-Bloch wave (FB wave) \cite{Russell1,brillouin}. It is characterized by the wave vector in the first Brillouin zone, referred to in the following as the fundamental wave vector \cite{fundamental}. We will discuss the characteristics of the FB wave in detail in Sec. IV. The intersections between EFS and construction lines represent the possible wave vector values for the FB propagating beam(s).  However, not all of these intersections correspond to a signal pointing away from the source (causal signal). So, for each wave vector intersection we need to determine the direction of the corresponding signal, which is nothing but the direction of the corresponding energy velocity. We will show in Sec. VII that the energy velocity $\bf v_{e}$ is equal to the group velocity $\bf v_{g}$ for the photonic crystal. Now, $\bf v_{g}$ is ${\bf \nabla_{k}} \omega$. Thus, the geometric properties of the gradient require the propagating signal to have a direction normal to the EFS at a certain wave vector point in ${\bf k}$-space and to point towards increasing frequencies $\omega$. Accordingly, before we proceed with a wave vector diagram analysis, additional knowledge for the sign of the product $\bf {\bf \nabla_{k}} \omega  \cdot k$ is necessary. Essentially,  we need to know the curvature of the relevant band.  
\par    
As already mentioned in the introduction, wave vector diagrams have been used before in the PC literature \cite{Notomi,Gralak,Joann,fotein,kosaka,kosaka2}, but were drawn in the first Brillouin zone only. Although in many cases this methodology suffices, for a complete treatment such diagrams should include the EFS shown in the repeated zone scheme.  A complete treatment yields all possible waves that can couple inside the photonic crystal. All intersections between the construction lines and the EFS in the repeated zone scheme should be accounted. Apparently, in our repeated zone treatment, some intersections may have wave vectors falling  outside the first BZ. These, should be folded back to the first zone, in order to obtain the fundamental wave vector of the FB wave. The folding process involves adding or subtracting an appropriate reciprocal lattice vector. Actually, in our study cases, where the interface is chosen along a symmetry direction,  only the primary construction line (the one with m=0) is sufficient. We refer to it, as simply construction line or ${\bf k}_{\|}$ conservation line. To convince the reader, we show an example (see Fig. 2) with all the construction lines present. However, for simplicity, in all the other diagrams the higher order construction lines are dropped. Summarizing, in any case we have fixed frequency (EFS contour), fixed sign $ \nabla_{\bf k} \omega \cdot {\bf k}$, and  fixed parallel component of wave vector (construction line). Our methodology stems from the properties of the FB wave, which we discuss in more detail in Sec. IV.  Following the folding process many points may fall onto the {\bf same} point in the first zone. We emphasize that in this case, all points give rise to one beam only. We refer to these points as ``equivalent'' points. Only different wave vector points in the first zone (after the folding process) yield different beams, provided the corresponding signal is causal. We distinguish between effects that stem from the first and the higher Brillouin zones. A beam that originates from $\bf k$ points in the first BZ is the ``transmitted'' beam, while we refer to beams coming from ${\bf k}$ points in the higher order zones, as higher order beams.   
\par 
The PC medium can also give rise to multiple reflected beams appearing in the medium of incidence. We can choose to determine these graphically. In this case, one would need to keep all construction lines determined by Eq. (1). Alternatively, the angle of an order m reflected beam, when the medium of incidence is air, is given by the following simple formula \cite{sakoda}:\\

\begin{eqnarray}
\theta_{m}=\sin^{-1}\frac{k_{\|,m}}{\omega /c},
\end{eqnarray}
provided \\ 
\begin{eqnarray}
k_{\|,m}^2<\omega^{2}/c^{2}.
\end{eqnarray}
(with $k_{\|,m}$ given by Eq. (1)). If condition (3) is not satisfied $\forall m\ne 0$, then no additional higher order reflected beams exist.  To avoid confusion with the higher order waves in the PC medium, we refer to the latter as higher order Bragg reflected beams. Notice that formula (2) also provides the angle for the order m outgoing beam, in the case that air succeeds the PC slab material.
\par  
 In Fig. 2 we show the wave vector diagram for the case of Fig. 1(a) drawn in the repeated zone scheme. Note that all the EFS are calculated with the use of the plane wave expansion method. We note that whenever we refer to the PWE method \cite{pwe1,villen,pwe2} for the H-polarization case, we applied Ho's method instead of the inverse expansion method, since the former is proven to show faster convergence \cite{villen,pwe2}. The bold green dot-dashed line in Fig. 2 represents the construction line. It intersects points A, B of the EFS in the first zone (black circle) and points A2, B2, A3, B3 of the EFSs in the higher order zones. We fold points A2, B2, A3, B3 back to the first zone by adding $\bf{G_{n}}$=n $\bf G_{oy}$ (see figure) (where n=-2 for points A2, B2 and n=+2 for A3 and B3). Notice they all fall back onto points A and B. The case of Fig. 2 corresponds to a band with negative curvature. Therefore, A has ${\bf v_{e}}$ pointing away from the source, while B has ${\bf v_{e}}$ pointing towards the source. This means that only point A contributes to a propagating beam, the ``transmitted'' beam. We indicate the propagating beam with the bold orange arrow in Fig. 2, and the bold solid arrow in the FDTD simulation in Fig. 1(a).  As we mentioned earlier, only the zeroth order construction line is really necessary in the determination of the propagating beams. However, in Fig. 2 for didactic reasons, we plotted the higher construction lines as well (dotted cyan vertical lines). Each is displaced in respect to the zeroth order one (bold green vertical line) by ${\bf G_{int}}$= $\lambda$ $\bf G_{ox}$ (where $\lambda=\pm 1, \pm 2$ e.t.c). We plotted in the figure only the higher order construction lines with $|\lambda|=1$ . We account for all intersections of these lines. Subsequently, we fold them back to the first zone by adding/subtracting an appropriate reciprocal lattice vector ($\bf G$=n $\bf G_{oy}$+$\lambda$ $\bf G_{ox}$ with $n\pm 1$, $\lambda \pm 1$). All fall back onto either A or B. Evidently, only the zeroth order construction line is sufficient. We checked and determined this generally holds for any two-dimensional PC slab, provided that the interface is cut along a symmetry direction. From now on we draw only the zeroth order construction line. In the insert of Fig. 2 we show a wave vector diagram for the same case, but drawn in the first zone only. Apparently, in this case an analysis in the first BZ gives identical results with an analysis with the modes drawn in the repeated zone scheme.
\par  
In Fig. 1(b) we see that two beams coexist (birefringence). This case corresponds to a band with positive curvature ($\bf v_{e} \cdot k>0$). A wave vector diagram in the repeated zone scheme is seen in Fig. 3. The green bold solid line in the diagram is the zeroth order construction line for this case. We mark the intersections of this line with the modes in all the zones and fold back those falling outside of the first zone (by adding $\bf G$= -$\bf G_{ox}$+n $\bf G_{oy}$ with $n=\pm 1$). Notice, points A2, B2, A3, B3 when folded back in the first zone, fall onto points A', B' that are different from A and B. From the set of wave vector points resulting in the first zone (A, B, A', B') only B and B' give a signal that propagates away from the source. The respective signals are indicated with the bold (point B) and dotted (point B') orange vector in Fig. 3. They correspond to the bold and dotted black arrows in the FDTD simulation of  Fig. 1(b). Actually, the first beam (solid arrow in Fig. 1(b)) can be explained with an analysis in the first zone (see insert of Fig. 3) and is the transmitted beam. Nevertheless, the beam that corresponds to the dotted arrow clearly stems from a higher order zone, i.e., a higher order effect. The latter would be impossible to predict with an analysis within the first Brillouin zone. 
\par
In Fig. 4 we see the EFS plotted in the repeated zone scheme for the case of Fig. 1(c). The construction line (bold green solid line) intersects point A and B in the first Brillouin zone and points A1, B1, A2, B2 in the higher order zones. These fall onto points A' and B' when they are folded back to the first zone (with a similar process as in Fig. 3). Taking into account the sign of $\bf v_{e} \cdot k$, in this case positive, from all intersections A, B, A', B' only B and B' yield a propagating beam (FB wave) (bold and dotted orange arrow in Fig. 4). The one that corresponds to point B can be determined with an analysis within the  first BZ (see insert) and is the transmitted beam. The second one is a higher order effect. The origin for  this higher order wave is similar to that of Fig. 3. However, because the EFS are anisotropic in this case, both beams (transmitted and higher order) are negatively refracted beams. These types of higher order waves, when EFS are anisotropic and broken, are unique to the triangular structure. They stem from the six-fold symmetry of the modes in the wave vector space. For square structures with equifrequency contours broken (4-fold symmetrical in this case), birefringent effects of this kind cannot be observed. 
\par
In Fig. 5 we show the wave vector diagrams in the repeated zone scheme that corresponds to the case of Fig. 1(d). The $\bf k_{\|}$ conservation line intersects several points (A2, B2, A3, B3). However all of the points, when folded back in the first zone, fall onto either point A' or B' (are equivalent to A' and B' respectively). Since $\bf v_{e} \cdot k >$0, only the wave vector at B gives a propagating FB wave, with energy velocity indicated by an orange solid arrow in the figure.  This is the sole propagating beam shown with a black dotted arrow in the corresponding FDTD simulation in Fig. 1(d) and is essentially a higher order beam. Note, an analysis within the first BZ (see insert of Fig. 5) predicts no propagating beam in this case. A wave vector analysis in the repeated zone is needed.
\par
Our preceding analysis shows that in any general 2D PC system, the direction(s) of the propagation beam(s) can always be determined with careful use of the wave vector diagrams in the repeated zone scheme. Notice the excellent agreement between the theoretical prediction ---derived from the wave vector diagrams--- {\hspace{1mm}} and the actual FDTD simulations seen in Fig. 1.  We do not discuss the aspect for the ``rightness'' of propagation in this section. We will see in Sec. IX that cases with $\bf v_{e} \cdot k <0$  represent a backwards (left-handed) wave. We note at this point that only the negatively refracted beam in Fig. 1(a) is a backwards wave. Actually, the mechanisms that lead to the formation of a negatively refracted beam in the cases of Figs. 1(a) through (d) are distinctly different. In particular, in Fig. 1(a) a negatively refracted beam is formed, because the perpendicular component of the wave vector reverses sign when meeting the air-PC interface. In fact, this is the same mechanism leading to a formation of a negatively refracted beam in the homogeneous negative index medium (NIM). However, this is not the case for Figs. 1 (b) through (d). Notice that in all these cases the perpendicular component of the wave vector does not reverse sign at the PC interface. For the case of Fig. 1(c), the negatively refracted beam , indicated with the solid arrow, is due to the anisotropy of the EFS in ${\bf k}$-space. The negatively refracted beams in Figs. 1(b) and 1(d) stem from modes in the higher order zones. In other words, both are a higher order effect. The same holds for the second negatively refracted beam in Fig. 1(c) (dotted line). Nevertheless, still the nature of the higher order waves in Figs. 1(b) and 1(d) (dotted lines) is different. In the case of Fig. 1(b), the component of the incident wave vector parallel to the interface ${\bf k_{\|}}$ falls within the limits of the surface 1D BZ. This zone extends between $-\pi/b_{str,cut},\pi/b_{str,cut}$, with $b_{str,cut}$ being the interface periodicity given by expression (A1). On the contrary, in the case of Fig. 1(d), ${\bf k_{\|}}$ falls outside the 1D surface BZ limits. In order to distinguish between the two waves, we refer to them as type I and type II, respectively. Type II higher order waves are present because of the periodicity across the interface. Type I higher order waves are present because of the periodicity of the whole  bulk photonic crystal. Hybrid higher order effects of type I and II can also be observed in some cases. Type II waves are generic and can be observed for any case, provided that frequency and/or angle of incidence is high enough. Type I higher order waves are particular to the specific lattice type, symmetry direction of interface, and EFS features.
\par   
We emphasize that higher order waves of type I always coexist with a beam deriving from the first BZ (transmitted beam). Thus, whenever a type I wave is present, birefringence is observed. This is the case of Figs. 1(b) and 1(c). Birefringent effects in photonic crystals were observed experimentally by Kosaka et  al. \cite{kosaka2}. In their study, for the frequency where the double branching of the beam was observed, two-band solutions exist. In other words, two different equifrequency contours exist within the first Brillouin zone for the relevant frequency. When multiple bands exist for  a certain frequency, the procedure we just described in detail must be repeated for each separate band. Many more beams can propagate in these cases. We note that Born and Wolf \cite{born} (as well as Yariv and Yeh \cite{yariv2}) adopted an effective medium approach to describe the 1D layered medium. They found that it effectively behaves as a homogeneous medium with optical anisotropy. Therefore, the 1D layered medium is capable of showing birefringent effects. They termed these effects as ``form'' birefringence to stress the fact that these originate from anisotropy on a much larger scale than the molecule. In optically anisotropic materials, the tensor property of the permittivity introduces two solutions for the dispersion relation (ordinary and extraordinary). The two different dispersion relations give two different equifrequency surfaces in the wave vector space and lead to the familiar birefringent phenomena in these media. In a way, we can say that multi-fringent phenomena in PC's arising from multiple bands appear for similar reasons as the ones in the optically anisotropic materials. In essence, multiple bands imply multiple dispersion relations for a certain frequency region, and, therefore, multiple EFS within the first BZ. However, all may be extraordinary (non-circular in 2D). In this paper, we observe in Figs. 1(b) and 1(c) birefringent effects that have a totally different origin. In these cases, there is a single band for the operation frequency, i.e., a single branch for the dispersion relation,and, therefore, a single EFS exists in the first BZ. The two beams arise because the modes, represented by the EFS, repeat themselves periodically in the wave vector {\bf k}-space. This means that the periodicity of the PC lattice comes into play in two different ways as far as birefringent effects are concerned. It introduces the possibility of having multiple dispersion relations within the first BZ for a certain frequency region, as in the case of Kosaka et al. \cite{kosaka2}. On the other hand, the periodicity causes the modes to repeat themselves in reciprocal space.  This is responsible for the beam doubling effects we observed in Figs. 1(b) and 1(c).  
\par  
The existence of multiple beams in the medium of incidence and the medium succeeding the PC slab, as well as how these relate to the transmission properties, were previously studied \cite{sakoda}. Despite these studies, the relation between the existence of multiple beams in the medium of incidence and inside the PC has not yet been carefully examined. Incidentally, Luo et al. \cite{Joann} state that the condition necessary to obtain single beam propagation inside the PC is $\omega \leqslant 0.5 \times 2 \pi c/a_{s}$ , where $a_{s}$ is the interface period. The quoted condition can be rewritten as $\tilde f \leqslant a/2b_{str,cut}$, with $b_{str,cut}$ given by (A1). In fact, if EM waves are incident in the PC slab from air, this condition guarantees the absence of any higher order Bragg reflected beams for {\bf any}  angle of incidence (see Appendix I(a)). For a triangular lattice cut along $\Gamma M$, this condition becomes $\tilde f \leqslant ~0.289$. However, FDTD simulation results that we present in the following show  this condition {\bf does not} guarantee single beam propagation inside the PC medium. We consider the case of dielectric rods with permittivity  $\epsilon=60.$ and radius r=0.37 a. This is a case qualitatively similar to that of  Fig. 1(b) (Fig. 3), but with a much lower relevant frequency ($\tilde f=0.275$). We stress that one band only corresponds to this operation frequency. Evidently $\tilde f=0.275$ is below the quoted limit, which means no higher order Bragg reflected beams exist for any angle of incidence. Indeed, in Fig. 6 we observed only one reflected beam. Notice, however, the clear presence of two propagating beams (solid and dotted arrow). The second beam (dotted arrow) is a higher order wave of type I, like the one shown in Fig. 1(b). 
\par 
A single beam propagation condition cannot be derived always in a simple manner and one should, in general, carefully examine the wave vector diagrams in the repeated zone scheme. When after the folding process, only a sole wave vector in the first BZ gives a causal FB wave, only then do we have single beam propagation. Consequently, the presence of only a single reflected beam is neither a prerequisite nor does it guarantee the presence of a single beam coupling into the PC medium. Also, note in Fig. 1(d) the clear presence of a higher order reflected beam, while there is only one propagating beam. For certain simple cases of isotropic EFS, we will discuss the conditions for single beam propagation in Sec. VI. 
\par
\vspace{1.5cm} 
{\bf IV. FLOQUET BLOCH WAVE AND PHASE VELOCITY}
\vspace{0.2cm}
\par
Consider the magnetic field of an H-polarized wave inside a two-dimensional periodic photonic crystal structure for the case of H (TE) -polarization.  
\begin{eqnarray}
  {\bf H}(r,t)= \frac{1}{\sqrt {A_{WS}}} e^{i{{\bf k} \cdot {\bf r}}} 
\sum_{\bf G}{H_{{\bf G}} ({\bf k},\omega_{n,{\bf k}}) e^{i{\bf G}\cdot{\bf r}}} \hspace{0.5mm} e^{-i\omega_{n,{\bf k}} t} \hspace{1mm} \hat z.
\end{eqnarray} 
\newline  
 $A_{WS}$ is the area of the Wigner-Seitz cell, $\hat z$ is the unit vector in the direction out of the plane of periodicity (i.e., the direction of the cylindrical rods) and {\bf G} is a reciprocal lattice vector. The coefficients $H_{{\bf G}}$ are determined from the eigenvalue equations obtained from the PWE method \cite{pwe1,villen,pwe2}. Apparently the above expression for the field satisfies the Floquet-Bloch theorem  \cite {brillouin} for a periodic medium. A field that propagates according to expression (4) is known as a Floquet-Bloch wave \cite{brillouin,Russell1} with {\bf k} lying in the first zone. Any attempt to express the propagation solution in terms of wave vectors $\bf k^{'}$ lying outside the first BZ results in an expression equivalent to Eq. (4) (see Appendix I(b)). So, the wave vector chosen in the first zone is what characterizes a propagating FB wave. We call this the fundamental wave vector \cite{Russell1,fundamental}. Hence the term, ``equivalent points,'' describing points in ${\bf k}$-space separated by a reciprocal lattice vector. This property of the FB wave explains the general recipe that we followed in the preceding section to determine the propagating waves. Clearly in the 2D periodic system, all plane wave components contributing to the FB wave with fundamental wave vector {\bf k} (expression (4)) propagate together, not separately, with a common energy velocity, $\bf v_{e}$. Note, no individual plane wave components serve as a separate solution of Maxwell's equations. As a result we do not see clear phase fronts, but rather have phase-like fronts with  a ``wiggly'' profile (see for example  the FDTD simulation presented in Fig. 1). This type of profile manifests a strong plane wave component mixing \cite{Notomi}, present in PC crystals with high refractive index modulation.
\par   
The questions of how one should approach the subject of defining a phase velocity for the FB wave is still unanswered. What would really be the physical meaning for such definition. Yariv \cite{yariv} defined a phase velocity for a 1D periodic system (see Appendix I(c)). Equivalently, for the two dimensional system this definition would be
\begin{eqnarray}
{\bf v_{p}}=\frac{\omega}{K_{o}^2} {\bf K_{o}},
\end{eqnarray}  
with ${\bf K_{o}}={\bf k}+{\bf G_{o}}$ being the plane wave component that has the larger amplitude ${H_{{\bf G_{o}}}}$ in expression (4). In other words, it is the wave vector of the predominant plane wave component. Many PC studies \cite{datta,yariv2,max_gar} focused on the long wave length limit, where such a definition would be appropriate \cite{yariv}. Nonetheless, the interesting refractive behavior of the photonic crystal reveals itself in the higher bands. Unavoidably, the subject of phase velocity in the photonic crystal requires some rethinking. As a matter of fact, Notomi \cite{Notomi}, as well as Kosaka \cite{kosaka3}, defined a phase index that corresponds to the fundamental wave vector within the first BZ, of the FB wave. Accordingly, the phase velocity would be:\\
\begin{eqnarray}
{\bf v_{p}}=\frac{\omega}{k^2} {\bf k},
\end{eqnarray}
where ${\bf k}$ is in the first BZ.\\
\par
There is an apparent contradiction between these two definitions as given by expressions (5) and (6), respectively. To investigate for the physical meaning of the phase velocity in a periodic medium, we will study numerically the field patterns of the propagating wave in the next section. 
\\
\par
{\bf V. NUMERICAL DETERMINATION OF THE PHASE VELOCITY}
\hspace{0.2cm} 
\par 
We consider two cases of almost isotropic EFS. In both cases the fundamental wave vector of the propagating FB wave is chosen to lie along a symmetry direction. We choose again the structure of Notomi (i.e., the same structure as in Fig. 1(a)) and two frequencies a) $\omega_{1}$=0.58 $2\pi c/a$ lying in a band with negative curvature and b) $\omega_{2}$=0.48 $2\pi c/a$ lying in a band with positive curvature. In order to extract information about the phase velocity in the system, we need to compare the time-independent fields at various points along the propagation direction. This methodology is analogous to the one followed by Ziolkowski and Heyman \cite{ziol}, where the negative phase index was numerically confirmed for a homogeneous slab material with $\epsilon=-1$ and $\mu =-1$.
\par  
For this purpose, we consider a pulsed signal $f(t) \cos(\omega_{0} t)$ with\\
\begin{displaymath}
f(t)=\left \{ \begin{array}{ll} 0 & \textrm{if $t<t1$ }\\
\frac{(t-t1)^2}{\alpha^2+(t-t1)^2} & \textrm{if $t1<t<t2$\hspace{1mm}. \hspace{5mm} (A2)}\\
\frac{(t-t3)^2}{\alpha^2+(t-t3)^2} & \textrm{if $t2<t<t3$}\\
0 & \textrm{if $t>t3$}
\end{array}\right.
\end{displaymath}
\par 
The parameters are chosen to give a broad  pulsed signal in time, with a small $\delta \omega$ around the  operation frequency $\omega_{0}$. For operation frequency $\omega_{1}$  the parameters are $\alpha$=21.9 T, t1=19.6 T, t2=499.7 T, and t3=979.9 T. For operation frequency $\omega_{2}$ they are $\alpha$=18.1 T, t1=16.2 T, t2=413.6 T, and t3=810.9 T. The period T of the EM wave  is different for the two cases and the parameters are chosen to correspond to the same actual time. The pulse is launched normally to the photonic crystal structure, with an interface cut along the $\Gamma K$ symmetry direction. We monitor the magnetic field {\bf H} for each time step for a long time, for certain points along the propagation direction. We refer to these points  as detector or sampling points. The Fourier transform of the time series ${\bf H(y,t)}$, where y the coordinate of a detector point, yields the corresponding amplitude ${\bf H(y,\omega)}$.  
\par           
Before we proceed with our analysis, we should mention that in order to make any assessment regarding the phase velocity the field patterns inside the slab should be as close as possible to the infinite system patterns, given by the FB wave expressed in (4). For this purpose, our structure should emulate a semi-infinite PC slab.  Any wave that couples into the slab will undergo multi-reflections between the two interfaces. In order to achieve our goal, i.e., a structure acting similar to a semi-infinite slab, we must somehow eliminate or reduce substantially the amplitude of any reflected beam originating from the second interface. For this purpose, we consider a periodic structure consisting of 30 sites along the lateral direction and 100 rows along the propagation direction. Of these 100 rows, 30 rows consist of rods  with dielectric constant 12.96 embedded in air. This part of the structure is the area of concentration. We take detector points that monitor the field as a function of time in this area. In the remaining 70 layers, we introduce absorption (both in the sites and in the background) so that the field is attenuated before exiting the crystal.  We introduce electric and magnetic conductivity ($\sigma_{e}$ and $\sigma_{m}$, respectively) to each numerical cell, so that the impedance of the each grid cell in the absorptive layer would be the same as the corresponding one in the non-absorptive layer. It is equal to $\sqrt{\frac{\mu_{0}}{\epsilon_{0} \epsilon_{i,j}}}$, where $\epsilon_{0}$ and $\mu_{0}$  are the vacuum permittivity and permeability, and $\epsilon_{i,j}$ is the relative permittivity of the 2D grid cell located at the point with grid coordinates (i, j). This is possible when the conductivities that we introduced follow the relations: $\sigma_{e}=\epsilon_{i,j} \sigma_{0}$ and $\sigma_{m}={\frac{\mu_{0}}{\epsilon_{0}}} \sigma_{0}$ , where $\sigma_{0}$ is a conductivity parameter. If the parameter $\sigma_{0}$ is chosen very low ($10^{-5} \Omega^{-1}m^{-1}$), the reflections of the beam when entering the absorptive layer are low. In order to make a rough estimate of the EM wave energy that gets reflected back to the non-absorptive layer where we monitor the fields, we look at the attenuation profile of the fields inside the absorptive layer. In addition, reflections can occur when the EM wave meets the absorptive boundary. To check these, we considered oblique incidence. Overall, we find the amplitude of the field that gets reflected back into the non-absorptive layer is about $10\%$ of the amptitude of the refracted EM wave.  
\par   
The set of points that serve as detectors are chosen normal to the surface, which coincides with the propagation direction $\hat y$, chosen along the $\Gamma M$ symmetry direction. Their respective locations  are a distance  $b=\sqrt{3} a$ apart, essentially the periodicity for the propagation direction. We take the Fourier transform of the time series representing the evolution of the magnetic field at a certain point \cite{numrecipe}. Afterwards,  we calculate the ratio's $H(\omega,d_{i+1})/H(\omega,d_{i})$. $\omega$ represents the frequency of the input pulse train and $d_{i}$ the location of the $i^{th}$ detector point. Since the distance between the detectors is one period along the propagation direction, this ratio should be equal to $\exp (i k b)$, with k restricted in the first BZ. Therefore, by studying the field patterns, we can extract information about the phase velocity, defined in accordance with Notomi \cite{Notomi} (expression (6)).
\par
We calculate the field ratio at adjacent detector points and extract the wave vector from the following formula\\
\begin{eqnarray} 
k_{i}= \frac{1}{i b} ln \frac{H(\omega,d_{i+1})}{H(\omega,d_{i})},
\end{eqnarray}
where i stands for the $i^{th}$ detector point. Notice that the Fourier transformed fields are complex and therefore the logarithmic function is complex and multivalued. We record all possible values with the real parts falling inside the first BZ. Taking the average for the various detector points for the case of $\omega= \omega1$, we find that two possible solutions exist for k: 1) k=$(1.47+0.005 i)\pm(0.01+0.008 i)\hspace{1mm} a^{-1} $ and 2) k=$(-2.16+0.005 i)\pm(0.01+0.008 i)\hspace{1mm} a^{-1}$. In order to choose the correct solution, we further study the field patterns. We also consider the ratio of two observation points located around the middle of the 30-cell PC layer, separated by $\Delta y=b/31$. The field ratio  determined by the FDTD simulation is  $3.05-0.63 i$. Now we calculate the same ratio theoretically with the PWE expansion method \cite{pwe1,villen,pwe2} for the two possible wave vector solutions determined from Eq. (7). Using solution (1) for the wave vector (real part only) in PWE we obtain a field ratio of $4.3+2.3 i$. For the second solution, we obtain a field ratio of $2.98-0.52 i $. Apparently only solution (2) gives a ratio that agrees well with the FDTD results. We note that in order to eliminate any discrepancies resulting merely from the discretization, we used in the PWE the actual numerical dielectric grid used in the FDTD. However, there is still a small discrepancy between the PWE field ratio and numerical FDTD ratio that may stem from a combination of the following ---angle span of incident source, Fourier transform zero padding errors, reflections from absorptive boundary layer, etc. This is the same reason for which a small imaginary part is present in the wave vector value. Also note there exists some ambiguity associated with the exact location of the first interface, since we are dealing with a periodic medium \cite{oldano}. In the boundary layer the field values may deviate from the values given by the FB wave expression (Eq. (4)). Therefore, we place the first detector point at the center of the second row of cylinders and assign $d_{1}=0$.
\par
In the analysis above we used the FDTD field patterns in the ``semi-infinite'' slab to determine that the wave vector inside the photonic crystal for the case with frequency $\omega= 0.58 \times 2 \pi c/a$ is $k=-2.16 \hspace{1mm} a^{-1}$. This value is in good agreement with the corresponding value from an EFS analysis ($k=-2.48 \hspace{1mm} a^{-1}$) \cite{discrepancy}. Expression (6), with the use of the wave vector obtained from the field pattern analysis, gives ${\bf v_{p}} \sim -1.69 \hspace{1mm} c \hspace{1mm} \hat y$ (c being the velocity of light). Following the same procedure but for the case with frequency  $\omega_{2}=0.48 \times 2 \pi c/a$, we calculate a phase velocity ${\bf v_{p}} \sim 2.23  \hspace{1mm} c \hspace{1mm} \hat y$.  In both cases, we determined the magnitude and the sign of the phase velocity. The field pattern analysis for a semi-infinite slab gives a negative phase velocity (i.e., opposite to the propagation direction) for the case with frequency ($\omega_{1}=0.58 \times 2 \pi c/a  $). This result implies that the ``rightness'' of the propagating beam is negative in this case. So, a field pattern analysis with the FDTD method for a semi-infinite slab confirms the results we obtained from the wedge simulation experiment \cite{fotein}. We will discuss more about the ``rightness'' of propagation in Sec. IX. 
\par 
 In order to visualize the physical meaning for the phase velocity defined in Eq. (6) we plot the imaginary part of the magnetic field $H(\omega)$ for various detector points. We show the results for both cases with frequencies $\omega_{1}$ and $\omega_{2}$  in Figs. 7(a) and 7(b), respectively (open circles). The solid line is $A \sin k y$ where k is the wave vector as determined from the field pattern analysis above and y is the distance from the first detector point. Also, we choose an additional set of detector points, closely spaced and around the middle of the 30-row PC layer. We indicate the imaginary part of the corresponding Fourier transformed field of these points with stars in Fig. 7.  We see that the solid sinusoidal line passes closely to the field values corresponding to the first set of detector points (circles). However, the field values for the second set of detector points (stars) deviate substantially from the sinusoidal line and show very high variations.
\par  
From the FB wave expression we obtain (see Appendix I(d)):
\begin{eqnarray}
<H({\bf r}={\bf R})>= e^{i ({\bf k\cdot R}-\omega t)}<H({\bf r}=0)>,
\end{eqnarray}
where ${\bf R}$ is a Bravais lattice vector and $<>$ the spatial average within the unit cell. All detectors points of the first set are Bravais lattice vectors along the $\Gamma M$ symmetry direction. Our numerical results shown in Fig. 7 and expression (8) suggest that the phase velocity describes how fast the phase of the EM wave travels from period to period in the PC lattice. However, information of how fast the phase travels between adjacent points cannot be determined. Thus, it is clear why it is necessary to fold the wave vector in the first zone and define the phase velocity as in expression (6). We will return to the same subject and the appropriateness of definition (5) when we discuss photonic crystals with low index modulation in Sec. X.  
\newline
\par
{\bf VI. EFFECTIVE PHASE INDEX}
\hspace{0.2cm}
\par
It is desirable to define an effective phase index that is correlated with the phase velocity as defined in Sec. IV with Eq. (6). Correspondingly,\\
\begin{eqnarray}
 {\bf v_{p}}=\frac{c}{|n_{p}|} \hat k.
\end{eqnarray}
The sign of $|n_{p}|$ is chosen to reflect the behavior left- or right-handed of the PC system \cite{fotein} and in accordance with the left-handed literature \cite{Veselago}. This definition for the index is consistent with the analysis by Notomi \cite{Notomi}. However, we have seen in Secs. II and III that the photonic crystal system  can be quite complicated and various higher order effects may arise under certain conditions. One must bear in mind all these effects when interpreting the effective index for the photonic crystal system. Next, we analyze what this index represents, as well as what properties can be inferred from this index. 
\par
Let's consider a Snell-like formula,\\
\begin{eqnarray}
\sin \theta_{inc}=n_{p} \sin \theta_{ref},
\end{eqnarray}
for EM waves incident from air, into a PC medium with phase index $n_{p}$ (in general depends on the refracted angle $\theta_{ref}$). Sometimes, it is assumed that Snell's formula gives the direction of the propagating wave. This is not true, because, in general, in the photonic crystal the direction of the refracted wave vector and the direction of the propagating signal do not coincide. We discussed this in Sec. III. Note again, that the direction of propagation is always the direction of the energy velocity ${\bf v_{e}}$. Nevertheless, provided that  certain conditions apply, there will be only a single refracted beam in the photonic crystal, which propagates with an angle given by Snell's formula (Eq. (10)). These conditions are the following:\newline 
\hspace{25mm}1) Interface cut along a symmetry direction of the crystal.\\
\hspace{25mm}2) Almost isotropic equifrequency contours.\\
\hspace{25mm}3) $\bf k_{inc\|}$ falling between $-\pi/b_{str,cut}$,...,$\pi/b_{str,cut}$ \\
\hspace{25mm}4) $|n_{p}| <1/(2 C_{str,cut} \tilde f)$ \\
with $b_{str,cut}$ given by (A1) and \\
\begin{displaymath}
 C_{str,cut}=\left \{ \begin{array}{ll} 1 & \textrm{for triangular cut along $\Gamma K$}\\
\sqrt 3 & \textrm{for triangular cut along $\Gamma M$\hspace{1mm}.  \hspace{5mm} (A3)}\\
\sqrt 2 & \textrm{for square cut along $\Gamma M$}\\
0 & \textrm{for square cut along $\Gamma X$}
\end{array}\right. 
\end{displaymath}
\par 
Cases with slanted interfaces are more complicated to analyze, and higher order waves are more likely to occur, hence the first condition. The second condition guarantees that ${\bf v_{e}}$ and ${\bf k}$  are about coaxial. This implies that the angle derived from Snell's formula represents the propagating angle. The third condition guarantees that the wave is not a higher order wave (specifically type II as described in Sec III). If the wave is a higher order wave of type II and the EFS are isotropic, it will propagate with an angle that may be opposite in sign to the sign of the effective index. Finally, the last condition, in combination with the first and second condition, guarantees the absence of higher order waves of type I for any angle of incidence, i.e., it guarantees single beam propagation. However, if we desire a single reflected beam as well, then the condition (see Appendix I(a)), 
\begin{eqnarray} 
\theta_{inc}<\theta_{lim}=\sin^{-1}(\frac{a}{\tilde f b_{str,cut}} -1),
\end{eqnarray}
 with $\tilde f< {a}/{b_{str,cut}}$, should also be observed. Notice that if $\tilde f$ exceeds the value of ${a}/{b_{str,cut}}$ then higher order reflected beams occur for any angle of incidence.
\par
Even in the absence of condition three, if the rest of the conditions are valid we still obtain single beam propagation. In  this case, a modified Snell-like formula can be used to determine the single refracted beam,\\
\begin{eqnarray}
\sin \theta_{incp}=n_{p} \sin \theta_{ref},
\end{eqnarray}
where 
\begin{eqnarray} 
\theta_{incp}=\sin^{-1} (\sin{\theta_{inc}}-\frac{m a}{\tilde f b_{str,cut}}),
\end{eqnarray}
and m chosen so that $|\sin \theta_{inc}\hspace{1mm}- ma/(\tilde{f} b_{str,cut})|<min\hspace{0.1mm}(1,a/(2 b_{str,cut} \tilde{f}))$.\\
\par
To summarize the purpose of defining an effective index is that it gives qualitative and/or quantitative insight into the PC properties  such as the magnitude and direction of the wave vector, the ``rightness'' of the medium conveyed in the sign of the index and, under certain conditions the energy velocity. However, caution should be taken by the use of such a phase index. It does not contain information about higher order beams that can couple inside the crystal or in the air medium (reflected beams) or both. A wave vector diagram type of analysis always offers a more complete treatment for the system. We also alert the reader that in no way should this phase index be used in Fresnel type formulas \cite{mathieu} to determine the transmission and reflection coefficients of an EM wave incident on the PC structure.
\par 
\vspace{1.5cm}
{\bf VII. ENERGY AND GROUP VELOCITY}
\vspace{0.2cm}
\par
We have seen in Ref. \cite{fotein} that the sign of the product $\bf v_{e} \cdot k$ serves as a theoretical prediction for the sign of the ``rightness'' for the PC system. Such theoretical predictions agree well with the FDTD wedge simulation results \cite{fotein}. We used the equality between the energy velocity and the group velocity to easily identify the frequency regions with negative ``rightness'' \cite{fotein}. These would be the regions that correspond to a band with negative curvature. In Ref. \cite{sakoda} the equality between group and energy velocity is shown for 3D periodic dielectric structures. It is important to show that such equality holds in our 2D photonic crystal as well. In order to show this,  we derive expressions for both the group velocity $\bf v_{g}$ and energy velocity $\bf v_{e}$.
\par
In the following we derive an expression for the energy velocity $\bf v_{e} $ for the PC system for the H-polarization case. We start with the definition of the energy velocity,
\begin{eqnarray}
{\bf v_{e}}=\frac{<{\bf S}>}{<U>},
\end{eqnarray} 
where the brackets $<>$ refer to the spatial average within the unit cell of the time averaged quantities. ${\bf S}$ is the Poynting  vector and U is the energy density.
Thus, we calculate the spatial average of the time averaged quantities for the Poynting vector {\bf S} and the energy density U.
\par
The Poynting vector  is defined as
\begin{eqnarray}
{\bf S}=\frac{c}{4 \pi}{\bf E_{r}} \times {\bf H_{r}},
\end{eqnarray}
where ${\bf E_{r}}$ is the real electric field lying  in the plane of periodicity and ${\bf H_{r}}$ is the real magnetic field along the cylinder axis. The time averaged Poynting vector, if ${\bf E}$ and ${\bf H}$ are the respective complex quantities of the fields, is given then  by the expression,
\begin{eqnarray}
{\bf S}=\frac{c}{8 \pi}{\bf E} \times {\bf H^*} ,
\end{eqnarray}
with 
\begin{eqnarray}
{\bf H}(r,t)= e^{i({{\bf k} \cdot {\bf  r}}-\omega_{n,{\bf k}} t)}\hspace{0.5mm} {\bf v}_{n,{\bf k}}
\end{eqnarray}
and
\begin{eqnarray}
{\bf E}(r,t)= e^{i({{\bf k}\cdot {\bf r}}-\omega_{n,{\bf k}} t)} \hspace{0.5mm} {\bf u}_{n,{\bf k}},
\end{eqnarray} 
where
\begin{eqnarray} 
{\bf v}_{n,{\bf k}}= \frac{1}{\sqrt{A_{ws}}}\sum_{\bf G} {{\bf H}_{{\bf G}} ({\bf k},\omega_{n,{\bf k}}) \hspace{0.5mm} e^{i{\bf G}\cdot{\bf r}}}
\end{eqnarray}
and 
\begin{eqnarray}
{\bf u}_{n,{\bf k}}= \frac{1}{\sqrt{A_{ws}}} \sum_{\bf G}{{\bf E}_{{\bf G}}({\bf k},\omega_{n,{\bf k}}) \hspace{0.5mm} e^{i{\bf G}\cdot{\bf r}}}.
\end{eqnarray}
Evidently the fields  ${\bf E}$ and ${\bf H}$ satisfy Bloch's theorem. Note that ${\bf H}_{{\bf G}}= H_{{\bf G}} \hspace{1mm} \hat z$. The eigenvectors $\bf E_{G}$ and $\bf H_{G}$ are determined from the PWE method \cite{pwe1,villen,pwe2}.  From (16), (17) and (18) we have 
\begin{eqnarray}
{\bf S}=\frac{c}{8 \pi} {\bf S}_{n,{\bf k}},
\end{eqnarray}
with
\begin{eqnarray}
{\bf S}_{n,{\bf k}}={\bf u}_{n,{\bf k}} \times {\bf v}^{*}_{n,{\bf k}}.
\end{eqnarray}
Maxwell's  equation for the magnetic field is
\begin{eqnarray}
\nabla \times {\bf H}=\frac{i \omega}{c} \epsilon({\bf r}) \hspace{1mm}{\bf E}.
\end{eqnarray}
By substituting Eqs. (17) and (18) into Eq. (23) and by taking the cross product with  ${\bf v}^{*}_{n,{\bf k}}$, we obtain
\begin{eqnarray}
{\bf S}_{n,{\bf k}}=\frac{c}{\omega \epsilon({\bf r})} (-i {\bf v}^{*}_{n,{\bf k}} \times \nabla \times {\bf v}_{n,{\bf k}}+
{\bf v}^{*}_{n,{\bf k}} \times {\bf k} \times {\bf v}_{n,{\bf k}}).
\end{eqnarray} 
{\bf k} is the fundamental wave vector of the FB wave lying in the first BZ.
Then using expressions (19) and (24) we get

\begin{eqnarray}
 {\bf S}_{n,{\bf k}}=\frac{c}{\omega A_{WSC}} \sum_{\bf G,G^{'},G^{''}} { ({\bf k}+{\bf G}) \hspace{1mm} \eta_{\bf G^{''}} H_{\bf G}({\bf k},\omega_{n,{\bf k}}) }\times \nonumber
\end{eqnarray}
\begin{eqnarray}
 H_{\bf G^{'}}({\bf k},\omega_{n,{\bf k}}) \hspace{1mm} e^{i ({\bf  G}+{\bf  G^{''}}-{\bf  G^{'}}) \cdot {\bf r}},
\end{eqnarray} where $\eta_{\bf G^{''}}$ are the Fourier expansion coefficients for the inverse of the dielectric function.
\par
Now, taking the spatial average of ${\bf S}_{n,{\bf k}}$ and using the delta function expression (see Eq.(80) in appendix II), we calculate the spatial average of the time averaged Poynting vector. So\\

\begin{eqnarray}
<{\bf S}> =\frac {c^2}{8 \pi \omega_{n,{\bf k}}} \sum_{\bf G,G^{'}} {({\bf k}+{\bf G}) \hspace{1mm} \eta_{\bf G^{'}}}
H_{\bf G} ({\bf k},\omega_{n,{\bf k}}) \hspace{1mm}H_{{\bf G}+{\bf G^{'}}} ({\bf k},\omega_{n,{\bf k}}).
\end{eqnarray}
\par
For the time averaged energy density we have
\begin{eqnarray}
U=\frac{1}{16 \pi} \hspace{1mm} (\epsilon ({\bf r}) Re({\bf E E^{*}})+Re({\bf H H^{*}})).
\end{eqnarray}
Taking the spatial average of the timed average quantity given in Eq. (13) and using the normalization conditions,
\begin{eqnarray}
\sum_{{\bf G},{\bf G'}} { \epsilon_{{\bf G}-{\bf G^{'}}} {\bf E}_{\bf G} \cdot {\bf E}_{\bf G^{'}}}=1
\end{eqnarray}
and 
\begin{eqnarray}
\sum_{\bf G} { H_{\bf G} H_{\bf G}}=1.
\end{eqnarray} We find for the spatial average of the timed averaged energy density that 
\begin{eqnarray}
<U>=\frac{1}{8 \pi}.
\end{eqnarray}
Finally, after index manipulation,
  
\begin{eqnarray}
{\bf v_{e}}=\frac{c^2}{\omega_{n,{\bf k}}} \sum_{{\bf G_{1}},{\bf G_{2}}} {({\bf k}+{\bf G_{1}}}) 
\eta_{{\bf G_{1}},{\bf G_{2}}}  \hspace{1mm} H_{\bf G_{1}} ({\bf k},\omega_{n,{\bf k}})  \hspace{1mm} H_{{\bf G_{2}}}  ({\bf k},\omega_{n,{\bf k}}).
\end{eqnarray}
\par
We also calculated the group velocity for this case with the use of the ${\bf k}\cdot{\bf p}$ \cite{kp_phot,Busch1,Busch2} perturbation method (see Appendix II). We found that the group velocity is given by the same expression. Thus, we showed the equality between the energy velocity and the group velocity for our 2D system. In fact, we also followed the same methodology for the case with the electric field aligned along the cylindrical rods (E- or TM-polarization). In this case too, we confirmed the equality between the group and energy velocity as well. The energy velocity for the E-polarization case is given by the following expression
\begin{eqnarray} 
{\bf v_{e}}=\frac{c^2}{\omega_{n,{\bf k}}} \sum_{{\bf G}} { ({\bf k}+{\bf G}) E_{{\bf G}}^2 ({\bf k},\omega_{n,{\bf k}})}.
\end{eqnarray}
Evidently, the energy velocity for our 2D photonic crystal can be calculated with the use of the formulas (31) and (32) for the H and E-polarization cases respectively. One needs to know the fundamental wave vector, the index of the band of interest, and the FB wave coefficients (${H}_{\bf G}$ and ${E}_{\bf G}$, respectively). The latter are determined easily from the PWE method \cite{pwe1,villen,pwe2}.
   
\par
\vspace{1.5cm}
{\bf VIII. GROUP REFRACTIVE INDEX}
\vspace{0.2cm}
\par
A group index $n_{g}$ can be defined \cite{fotein} in accordance with traditional waveguide and optical fiber literature \cite{fiber}\\
\begin{eqnarray}
|n_{g}|= \frac{c}{|{\bf v_{g}}|}.\\
\end{eqnarray}
For a  PC structure with the same parameters as the one in Fig. 1(a), we calculated the group velocity vector for the $5^{th}$ band (band with negative curvature) and the $4^{th}$ band (band with positive curvature) for a range of frequencies where the EFS contours are ``almost'' isotropic. We used the ${\bf k}\cdot {\bf p}$ perturbation method \cite{kp_phot,Busch1,Busch2} result given in expression (82) of App. II a (which is equivalent to expression (31)). The results for the magnitude of the group velocity are shown in Fig. 8 for both bands and both symmetry directions ($\Gamma M$ and $\Gamma K$). Notice that the closer to the band edge, the better the agreement between the group velocities for the two symmetry directions. This is expected, since the degree of anisotropy increases as one moves away from the band edge. Alternatively, we can consider the PC as  an isotropic system with effective dispersive phase index $n_{p}(\omega)$. In this case,
\begin{eqnarray} 
|{\bf v_{g}}|=\frac{c}{|n_{g}|},
\end{eqnarray}
with,\\
\begin{eqnarray} 
n_{g}=|n_{p}|+\omega \frac{d |n_{p}|}{d \omega}.
\end{eqnarray}
\par 
 
We also show in Fig. 8 the results obtained from formulas (35)-(36) for comparison. Because of the small anisotropy in the EFS shape, we use for $n_{p}$ the average value of the two symmetry directions. For both bands the results given from Eq. (36) are shown as a solid line with circles. We see that the results from expression (36)  are in very  good agreement with the corresponding ones calculated from the ${\bf k \cdot p}$ perturbation method. This is especially true for frequencies very close to the band edge. So, in the cases that conditions 1-4 of Sec. VI are satisfied, expression (35) (with the use of (36)) provides a good estimate for the group/energy velocity of the single transmitted  beam. 
\par
 The sign of the group index manifests the sign of refraction at the air-PC interface. As in the case of the phase index $n_{p}$ though, caution must be exercised with the use and interpretations of the group index. We stress that the sign ${n_{g}}$ relates only to the sign of refraction for the transmitted beam. It does not contain any information for higher order waves of any of the two types discussed in Sec. III. 
 \par
\vspace{1.5cm}
{\bf IX. THE ``RIGHTNESS' OF THE PC SYSTEM: DESIGNING THE WEDGE-TYPE EXPERIMENT}
\par
\vspace{0.2cm}
We observed different negative refraction effects 
shown in Fig. 1 in Sec. I. It is important to be able to characterize the 
nature of propagation (left-handed or not) for the finite PC structure.
In Ref. \cite{fotein} we found that the presence of a negatively refracted beam does not necessarily
imply left-handed behavior. We confirmed \cite{fotein} that indeed the sign of the ``rightness'' follows 
the theoretical prediction for the sign of $n_{p}$ from the band structure. A wedge type of experiment that can determine unambiguously the PC's ``rightness'' can be designed in most cases. However, in Sec. III we observed different higher order effects that can potentially complicate the interpretation of such an experiment. It is important to take into consideration the symmetry properties of the crystal, as they reveal themselves in the cuts of the interfaces and in the eigenmodes in ${\bf k}$-space for the frequency of operation. In the following, we go over the intricate details of the wedge design. We start our analysis with the hexagonal PC, where we identify two general classes: those with ``almost'' isotropic EFS and those with anisotropic EFS (broken curves with 6-fold symmetry) in ${\bf k}$-space. Note that limiting cases between the two classes mentioned above exist. However, their respective  frequency range  is generally quite small and we will not treat such cases. It becomes evident that some a priori general knowledge for the system is necessary before designing and/or interpreting the wedge simulation/experiment. This knowledge can be extracted from the band structure. One locates ``almost isotropic cases'' at the band edge, where the band structure is bell-shape like. Away from the band edge, anisotropic EFS are  expected, which lie at the edge of the Brillouin zone. It is necessary to  know a priori in what general category the shape of the EFS falls into. In fact, in some cases an estimate for the magnitude of the transmitted wave vector may be needed as well. This knowledge can be obtained from the plane wave expansion method for the infinite system.  We stress though, that in any case the ``rightness'' for the PC from such a simulation/experiment is determined independently and no information regarding this quantity is borrowed from the infinite system analysis.
\par
We first consider a case with an almost isotropic EFS for the hexagonal lattices. An example is the case in Fig. 3 of  Ref. \cite{fotein} which corresponds to the system shown in Fig. 1(a) of this paper. Another example is the case of Fig. 1(b), that we choose to  describe in this section. In Fig. 1(b) we observed the coexistence of a negatively and a positively refracted beam when the wave refracts on the PC interface. The EFS for such a case in the extended zone are shown again in Fig. 9. 
The direction for both interfaces, first and wedged, are indicated with the turquoise lines. We chose $\Gamma K$ as the symmetry direction for both the interfaces. From now on we will refer to a certain wedge design as (sym1)-(sym2) where sym1 is the symmetry direction of the first interface, while sym2 is the symmetry directions of the second interface. So, in this case we consider a $\Gamma K$-$\Gamma K$ wedge design. Notice that the wedge separates the space into three different areas. The area where the fields come from (area 1), the area inside the wedge (area 2), and the area after the fields experience scattering (area 3). In Fig. 9 we draw the transmitted vector inside area 2 as it would be for a system with  ${\bf v_{e} \cdot k}$ positive, which, by the way, is the theoretical prediction. The light is normal to the first interface. According to the analysis in Sec. III, it is fairly obvious that there will be only one transmitted wave vector (indicated in Fig. 9 with the blue arrow). Naturally, in order to determine the outgoing beam we must apply the ${\bf k_{\|}}$ conservation at the wedged interface. We also checked in this case, as we did in Fig. 2, that only the zeroth order construction line (blue bold line in Fig. 9) suffices. We see this line intersects points P1, P2 in the first zone and points E1, E2, ..., E6 in the higher order zone. However, all points (E1, E2, ..., E6), when folded back to the first zone, fall onto either P1 or P2. In other words, they  are equivalent to points P1 and P2, respectively. Consequently, they do not give rise to additional $\bf k_{\|}$ values that would cause additional beams in area 3. This is true for any $\Gamma K$-$\Gamma K$ design, provided that $|n_{p}|<1/(2\tilde f)$. We note this condition is satisfied by the vast majority of cases with isotropic EFS. Our subsequent analysis concerns these cases. The fact that all intersections {E1, E2, ..., E6} are equivalent points comes as a virtue of the chosen symmetry direction for the wedged interface ($\Gamma K$).  We bring to the readers attention that this is not, in general, the case for any design. In particular,  $\Gamma K$-$\Gamma M$ or $\Gamma M$-$\Gamma M$ designs and for typical cases in this category, the $k_{\|}$ construction line on the wedged interface intersects points in the higher order zone not equivalent to the intersections in the first zone. However, we notice these intersections, when folded back to the first zone, yield a $k_{\|}$ value equal to $k_{t} \hspace{0.1mm} \cos \theta_{des} + 2 \pi m/b_{str,cut}$ with m equal to -1 and $b_{str,cut}$ given by (A1). The angle $\theta_{des}$ is different for different designs. It has the value of $60^{0}$ for the $\Gamma K-\Gamma M$  or $\Gamma M-\Gamma K$ designs, and $\theta_{des}=30^{0}$ for the $\Gamma K-\Gamma K$ and $\Gamma M-\Gamma M$ designs. Consequently, in all forementioned designs, all possible ${\bf k}_{\|}$ values are predicted by the Bragg formula. Accordingly, their outgoing angle is \\\
\begin{eqnarray}
\theta_{out,m}=\tan^{-1} \frac{k_{t}  \hspace{0.1mm} \cos \theta_{des} +\frac{2 \pi m}{b_{str,cut}}}{k_{\bot, m}},\end{eqnarray} 
for m that satisfy the condition,\\
\begin{eqnarray} 
{k_{\bot,m}^2}=\frac{\omega^2}{c^2}-(k_{t} \hspace{0.1mm} \cos \theta_{des} +\frac{2 \pi m}{b_{str,cut}})^2>0.
\end{eqnarray}
We alert the reader that only for a design with interfaces along symmetry directions can such simple formulas be implemented. If a design is considered with either one or both interfaces cut along a direction that does not coincide with a symmetry direction, then many more outgoing beams should be expected. 
\par 
Notice that $k_{t}$ is positive when parallel to the +y-direction and negative when anti-parallel to the +y direction. Apparently, from formula (37), the sign of the zeroth order outgoing beam coincides with the sign of $k_{t}$. Consequently, it is the position of the zeroth order outgoing beam that determines the ``rightness'' of the PC. Since the presence of more than one beam in area 3 may complicate the interpretation of the results, it is desirable to work with a design that can avoid these. In fact, for designs $\Gamma K$-$\Gamma K$ and $\Gamma M$-$\Gamma K$, a single beam in area 3 will be present for frequencies below $\tilde f_{0} \sim0.56$. From these two designs it is better to choose the $\Gamma K$-$\Gamma K$. In higher bands it is very likely a part of the energy to get reflected back inside the PC. The result of these consequent multi-reflections may introduce an additional beam. If we choose a $60^0$ wedge, the multi-reflected beam will always be along the normal to the wedge. Therefore, it is easily identifiable and should be ignored when present. This is the reason that the  $\Gamma K-\Gamma K$ should be used to study cases in this category. We performed a FDTD simulation using this design  for the system of Fig. 9. The results are shown in Fig. 10. We observe one outgoing beam in the positive hemisphere. Thus, the theoretical prediction that the PC is ``right-handed'' is confirmed. The second beam  around the normal is just the result of multi-reflections. As we mentioned above, if the frequency exceeds 0.56, additional beam(s) may be present in area 3. Very high frequencies for which multiple m's can satisfy Eq. (38) should be avoided in experimental studies. We suggest, as an upper frequency limit, the value of $\tilde f=0.75$. For frequencies below this limit, when using the wedge design indicated above and typical cases in this category,  the first order Bragg wave has an outgoing angle still quite  different in magnitude from the  $0^{th}$ order outgoing beam. Therefore, we can clearly identify the position of the latter in area 3. 
\par
 We proceed in our discussion with cases that have anisotropic EFS, i.e., broken curves with 6-fold symmetry in {\bf k}-space. Cases that fall in this category are more complicated and extra caution must be exercised in designing and interpreting the wedge type simulation results. An example for these cases, is that of Fig. 4 in Ref. \cite{fotein}, which is the same case as the one shown in Fig. 1(c) of Sec. II of this manuscript. We choose to analyze another example belonging to this category.  We consider rods with dielectric constant 12.96 and radius 0.30a in air, magnetic field along the cylinders (H-polarization), and operation frequency of $\tilde f=0.50$. Note, these cases have propagation modes only along $\Gamma K$. One can see in Fig. 4, for example, that  a horizontal line across $\Gamma M$ does not intersect any modes, neither in the first nor in the higher zones. This leaves us with two possible designs to work with: $\Gamma M$- $\Gamma M$ and $\Gamma M$- $\Gamma K$ design. Between the two we choose the first one, for reasons that we discuss later in this section. With the help of a wave vector analysis, it is easy to see that when the first interface is crossed, three different beams, having three different wave vectors, couple into the PC (area 2). One is the transmitted ${\bf k_{t}}$ (blue bold arrow) in Fig. 11, while the other two are higher order beams (${\bf k_{d1}}$, ${\bf k_{d2}}$). We draw the wave vectors in Figs. 11(a) as they would be for a ``right-handed'' PC and in 11(b) as they would be for a ``left-handed'' PC. At the wedged interface I2, we draw a line representing ${\bf k_{\|}}$ for each of them. The blue line represents the  ${\bf k_{\|}}$ value of ${\bf k_{t}}$, and the green dotted lines represent the  ${\bf k_{\|}}$ values  of ${\bf k_{d1}}$ and ${\bf k_{d2}}$. However, in order to determine the outgoing angles in area 3, corresponding to the the three wave vectors in area 2, a simple phase matching condition represented by the three lines in Fig. 11, is not sufficient. To predict all possible outgoing beams for each of the three ${\bf k_{\|}}$ values in area 2 (${\bf k_{t \|}},{\bf k_{d1 \|}},{\bf k_{d2 \|}}$), we must perform an analysis in the repeated zone scheme, as we did in Fig. 9. In fact, we should include all construction lines for each ${\bf k_{\|}}$ value to make sure that no possible outgoing beam is omitted. This is quite an elaborate process and we will not go over all the details. From this process, we observed, as a virtue of the symmetry the possible ${\bf k_{\|}}$ values at the interface I2 (${k_{\|,m,i}}$), are given by the Bragg formulas corresponding to each of the three wave vectors (${\bf k_{t}},{\bf k_{d1}},{\bf k_{d2}}$). Accordingly, 
\begin{eqnarray}
k_{\|,m,i}=k_{\|,i} +\frac{2 \pi m}{b_{str,cut}},
\end{eqnarray}
where the index i, represents the three possible wave vectors ${\bf k_{t}}$ ${\bf k_{d1}}$ and  ${\bf k_{d2}}$, and $b_{str,cut}=\sqrt{3} a$. Again, we restrict ourselves, at the lower frequencies, where at most the first order Bragg waves (for m=-1) can couple. An interesting observation is  the first order Bragg wave corresponding to ${\bf k_{d1}}$ has  ${\bf k_{\|}}$ value equal to  ${\bf k_{d2\|}}$ and vice versa. In fact, this result is due to the hexagonal lattice symmetry. So, in Fig. 11 we need only to add one more ${\bf k_{\|}}$ value (blue dotted line), that corresponds to the first order Bragg wave related to ${\bf k_{t \|}}$. The red circle in Fig. 11 represents the EFS in air. We indicate in Fig. 11 all possible outgoing beams for both cases of ``right-handed'' in (a) and ``left-handed'' in (b) photonic crystal. The bold solid and dotted blue arrows, and the bold dotted green arrows represent the directions of all four outgoing beams. Notice that the outgoing angles in each side of the normal to interface I2 are close. In addition, their values are quite different than the respective one in the opposite hemisphere. This is generally true for  typical cases in this category, if we restrict ourselves to frequencies below $\sim 0.60$. Therefore, cases with frequencies exceeding  $\sim 0.60$ should be avoided. Figure 11 indicates, that the position of the larger in magnitude angles determines the ``rightness'' of the PC system. In Fig. 12 we show the corresponding simulation for the system we chose as an example. We observe that the outgoing beam, with the larger in magnitude angle, lies in the negative hemisphere. This means that the PC is ``left-handed'' in this case. This agrees with the sign of ${\bf v_{e}} \cdot {\bf k}$ obtained from the band structure. Since anisotropic modes reside at the edge of the Brillouin zone, the magnitude of $\bf k_{t}$ will be quite large. Therefore, if the frequency is low,  $\bf k_{t\|}$ may not yield an outgoing beam (total internal reflection). Cases with frequencies below $\sim 0.50$ should also be avoided for study. These give total internal reflection even for small comparatively $k_{t}s$. If one draws Fig. 11 with a $30^{0}$ wedge ($\Gamma M$-$\Gamma K$ design), we would see that we obtain outgoing beams with comparable angles in the two hemispheres. Thus, interpretation of the results is vague for this design, which led us to use the  $\Gamma M$-$\Gamma M$ design.
 
\par
We focused our analysis above for the hexagonal lattices. In the following paragraph we discuss a corresponding appropriate wedge design for the square structures. Our criteria are the same as in the analysis of the hexagonal structures.  Then, given the choice, one should always prefer the interface cut for the wedged interface with the smaller periodicity, as Bragg waves (see formula 37) will begin to appear in  such cases for larger frequencies in area 3. Taking this into consideration and studying the modes in the extended zone for the square lattice, we have determined that for both classes of cases (isotropic and anisotropic) the appropriate wedge design is $\Gamma M$-$\Gamma X$. We note that the anisotropic cases for the square lattice have four-fold symmetry instead \cite{Joann}. The four-fold symmetry does not create higher order beams of type I, so only one ---not three as in the hexagonal case--- wave vector couples into area 2. In addition, the wedged interface is $\Gamma X$ in both cases and has periodicity a (lattice constant). For the anisotropic cases and square lattice PC wedge,  possibly we can have only one outgoing beam in area 3. Typical cases may also suffer from total internal reflection, if $\tilde f$ smaller than $\sim 0.45$. We note there may be ``almost'' isotropic cases that suffer from internal reflection. These will be cases that have an effective phase index with magnitude larger than $2/\sqrt 3$ for a  hexagonal  $\Gamma K$-$\Gamma K$ PC wedge and larger than $2/\sqrt 2$ for a square $\Gamma M$-$\Gamma X$ PC wedge. However, both of these cases are very rare (especially in the higher bands), since ``isotropic'' modes do not reside close to the edges of the Brillouin zone.
\par 
We note that our analysis applies for frequency regions where only a single band solution exists and $\omega({\bf k})$ is monotonic. In cases where either of these do not hold, multiple beams of different ``rightness'' may be present.
 
\vspace{1.5cm} 
{\bf X. HIGH INDEX VS. LOW INDEX MODULATION}
\vspace{0.2cm}
\par  
In the previous sections we focused our analysis in describing the propagation properties of EM waves for 2D crystals with high-index contrast between the constituents dielectrics. We have seen anomalous refractive effects that include birefringence. We provided a consistent recipe  based on the wave vector diagram and band structure properties of the system, which determines all properties of each propagating beam such as refracted angle, phase, energy velocity and ``rightness.'' Since the analysis in the preceding section focuses on PC's with high index modulation, it is important to investigate the limits of validity of our theoretical analysis. We will now examine photonic crystals with low index modulation in the context of all the properties that characterize the propagating beam discussed in the previous sections.
\par
We consider a 2D photonic crystal lattice that consists of dielectric rods in air with radius 0.35 a in triangular arrangement for the H-polarization case. We let the dielectric constant of the rods vary, starting from the value of 1.05 (value close to the dielectric constant of air), and investigate the photonic crystal's response as the dielectric constant of the rods increases. Unavoidably for the low index contrast cases it is not possible to isolate cases where there is only one band solution for a certain frequency. For the five different cases that we examine in our comparative analysis, we choose the operation frequency to lie approximately in the middle of the spectrum that corresponds to the $2^{nd}$ and $3^{rd}$ band. The cases that we analyze are the following, a) with dielectric constant $\epsilon=1.05$, b) with $\epsilon=1.2$, c) with $\epsilon=1.5$, d) with $\epsilon=2.0$, and e) with $\epsilon=5.0$. The operation operation $\tilde f=fa/c$ is 0.80, 0.78, 0.75, 0.70, and 0.54 for the cases (a) through (e), respectively. We first consider the case with $\Gamma K$ as the symmetry direction of the interface. The fields from the FDTD simulation for oblique incidence with angle $8^{0}$ with the surface normal are shown in Fig. 13.
\par
We notice that in case (a) the wave enters essentially undisturbed inside the PC with the angle of propagation pretty much the same as the angle of incidence. As the index contrast increases, the propagating  angle still remains close to $8^{0}$, but ``wiggly'' features start to appear in the phase fronts. Refraction angle remains positive. For index contrast 5.0 (case (e)) we see a beam in the negative direction. We have plotted the EFS for all five cases in the first zone in Fig. 14. In Fig. 15 we show the band structure for the two limiting cases ((a) and (e)). The solid lines correspond to the second band, while the dashed lines correspond  to the third band. In Fig. 14 the EFS that are closed curves belong to the second band, while the ones that are broken with a six-fold symmetry belong to the third band. Notice that the second band EFS are highly anisotropic for case (a) and become more and more isotropic as the index contrast increases. They become eventually ``almost'' circular for case (e). If we check the FB wave components in expression (4), we see that for cases (a) through (c) there is mainly one predominant FB wave coefficient. Mixing starts to appear in case (d) and becomes stronger in case (e).

Suppose that we could describe our periodic system as an effective medium that has a dielectric constant given by the average value of its components and, therefore, an effective index $n_{eff}$.\\
\begin{eqnarray}
n_{eff}=\sqrt{\epsilon*f+(1-f)}
\end{eqnarray}\\
with f being the filling ration and $\epsilon$ the rod dielectric constant. We note at this point that an $n_{eff}$ corresponding to Maxwell-Garnett theory \cite{max_gar} yields a similar propagating angle to the one predicted by the $n_{eff}$ of Eq. (40). The field solution will then be a plane wave, therefore:\\
\begin{eqnarray}
H({\bf r},t) \sim A e^{ i ({\bf K} \cdot {\bf r} - \omega t)}.
\end{eqnarray}
with $\bf K$ having the value that corresponds to a homogeneous medium with refractive index  $n_{eff}$ i.e., $K=n_{eff} \omega /c$. 
\par
Alternatively, one can consider the system as a periodic medium,  with the field solution an FB wave given by expression (4). For cases (a) through (c)  in the sum there is only one predominant term. Therefore,
\par
\begin{eqnarray} 
H({\bf r},t)=A e^{ i ({\bf k_{pred}} \cdot {\bf r}- \omega t)}+O(\epsilon ^2).
\end{eqnarray}
\par    
 We found that ${\bf k_{pred}}$=${\bf k+G_{0}}$, $\bf k$ in the first BZ and ${\bf G_{0}=}4\pi/\sqrt 3 \hspace{1mm} a^{-1}$ $\hat y$, with y the propagation direction ($\Gamma M$ in this case) ($x$ represents the lateral direction). In Table I, we show for all cases (a) through (e), the wave vector ${\bf K}$, assuming an effective medium with index given by Eq. (40) and the wave vector of the predominant component ${\bf k_{pred}}$ of the FB sum, that we determined from the PWE method. For cases (d) and (e) there are more than one plane wave components in the FB sum with significant magnitude. In these cases, for ${\bf k_{pred}}$, we chose the one with the larger amplitude. In Table I, we also show the angle of propagation $\Theta$, if we treat the system as a homogeneous system with index $n_{eff}$. Moreover, we show the angle of propagation $\theta_{pred}$ assuming that in the PC the predominant plane wave component propagates by itself. We notice a good agreement between ${\bf K}$ and ${\bf k_{pred}}$. A small discrepancy, becomes larger as the index contrast increases and mixing becomes stronger.  We observe that in the cases with almost no mixing ((a) through (c)), the angles ${\Theta}$ and ${\theta_{pred}}$ agree with the refracted angles observed in the simulations in Figs. 13(a) through (c). Nevertheless, in case (e) where mixing is quite strong, although ${\Theta}$ and  ${\theta_{pred}}$ are close, both are quite different from the actual refracted angle seen in the simulation  (Fig. 13(e)). The refracted angle in case (e) is negative.  This can be understood if we look at the energy velocity expressions in Sec VII. For cases (a) through (c), there is only one surviving term in expression (31) for ${\bf G_{1}}$ and ${\bf G_{2}}={\bf G_{0}}$. Therefore, the energy velocity becomes,
\begin{eqnarray}
{\bf v_{e}} \cong \frac{c^2}{\omega_{n,{\bf k}}} \hspace{1mm} ({\bf k}+{\bf G_{0}})  \hspace{1mm} \eta_{{\bf G_{0}},{\bf G_{0}}}  \hspace{1mm} H_{\bf G_{0}}^{2} ({\bf k},\omega_{n,{\bf k}})  \propto {\bf k_{pred}}. 
\end{eqnarray}

\begin{center}
TABLE I \\   
\vspace{5mm}       
\begin{tabular}{|c|c|c|c|c|c|c|c|c|}
\hline
$ {\bf }$& $K_{x}\hspace{0.2mm} $ & $ K_{y} \hspace{0.1mm} $ & $ |\bf K| 
\hspace{0.1mm}$ & $\Theta \hspace{0.1mm}$ & $ k_{predx} \hspace{0.1mm}$ & 
$ k_{predy} \hspace{0.1mm}$ & $ |\bf k_{pred}| \hspace{0.1mm}$ & $ 
\theta_{pred} \hspace{0.1mm}$\\ 
{\bf } & $(\pi/a)$ & $(\pi/a)$ & $(\pi/a)$ & $(^{0})$ & $(\pi/a)$ & 
$(\pi/a)$ & $(\pi/a)$ & $(^{0})$\\
\hline \hline  
(a) & 0.223 & 1.602 &  1.618 &  7.91 & 0.223 & 1.602 & 1.617 & 7.91\\
(b) & 0.217 & 1.613 &1.628 &  7.66 & 0.217 & 1.610 & 1.624  & 7.68\\
(c) & 0.209 & 1.645 & 1.658 & 7.23 & 0.209 &  1.632 &  1.645 & 7.29\\
(d) & 0.195 & 1.671 & 1.683 &  6.65 & 0.195 &  1.638 &  1.650 &  6.78\\
(e) & 0.150 & 1.794 & 1.800 &  4.79 & 0.150 & 1.835 &  1.841 &  4.68\\
\hline

\end{tabular}
\end{center}

\par 
\vspace{1cm}
So, when the index contrast is very low (cases (a) through (c)), only one component has a significant magnitude in the FB sum. Then, the direction of the energy velocity is very close to the direction of the predominant wave vector. However, as the mixing becomes stronger, the directions of the predominant wave vector and energy velocity can be very different (case (e)). Notice, in the latter cases, other wave vectors contributing to the FB sum can have an amplitude quite close to the predominant one. Alternatively, if we would like to describe the system graphically, for cases (a) through (c), we could accomplish this in two different ways. We could treat the system as a homogeneous system and draw its EFS as a circle in wave vector space with radius equal to $K=n_{eff} \omega/c$. But, we can consider also the medium as a periodic medium and follow the recipe of Sec. III. Both treatments in these cases give almost the same angle of propagation. This is in excellent agreement with what was observed in the FDTD simulation.  
\par
One might be tempted to describe a photonic crystal medium for cases with a low index contrast as a homogeneous medium with an index given by Eq. (40). However, the results we present in the following suggest that such a treatment would be erroneous. In fact, consider the five different cases of Fig. 13. In this case we take the same angle of incidence, but choose the interface along $\Gamma M$ and, therefore, the propagation direction $y$ along $\Gamma K$. We present the results in Fig. 16. Contrary to one's expectations for a homogeneous  medium with index $n_{eff}$, even for an index contrast as low as 1.2:1.0 (case b), we see three distinct refracted beams. These beams have propagating angles in excellent agreement with the predictions of a wave vector type of analysis in the repeated zone scheme. Therefore we can infer that the wave ``sees'' the periodicity of the medium even when the index contrast is low. The periodicity may introduce multiple bands for a certain frequency (see Fig. 15). In addition, because the system is periodic, the modes repeat themselves in wave vector space. Our results indicate, that the periodicity of the system should be taken into account even for the low index contrast cases. An effective medium approach may give inadequate results (predicts a beam close to the position of the middle beam in cases of Fig. 16(b) through (e)) or, in many cases, totally inaccurate for the high index contrast cases. For example, for case of Fig. 13(e), it predicts a positive instead of a negative refracted beam. 
For infinitesimally small index modulation, however, a wave vector analysis  may predict additional beams. In such a case the periodicity and band folding are a mere artificiality and the medium is essentially homogeneous and should be treated as such.
\par 
We notice also, that in Fig. 16, for cases (b) through (c),  although three beams are present, each has almost clear wavefronts. Actually we checked the field solution with the PWE method and we found that the FB wave describing each of these beams consists of only one predominant coefficient. Therefore, in these cases for each beam, we have information of how fast the phase given by ${\bf k\cdot   r}$ travels from point to point in space. So, in such cases, Yariv's picture (definition (5)) \cite{yariv} is appropriate for the phase velocity. Therefore, it is natural to ask, when does Yariv's picture begin to fall apart? To answer this, we consider a similar numerical experiment as in Sec. V. For detector/sampling points, we use adjacent points in the numerical grid space. We take normal incidence and $\Gamma M$ as the propagation direction y. Let $y_{i}$ represent the location of a point in the numerical grid space where we sample the field. We calculate the Fourier transformed field ratio at adjacent points, ${H(\omega,y_{i+1})}/{H(\omega,y_{i})}$. From this ratio, we extract ${\bf k}$ which will be
\par
\begin{eqnarray}
{\bf  k}=\frac{1}{i \Delta y} ln(\frac{H(\omega,y_{i+1})}{H(\omega,y_{i})}) \hat y,  
\end{eqnarray}
where $\Delta y=y_{i+1}-y_{i}={\sqrt 3}/{62}\hspace{0.2mm} a$, is the size of the numerical grid along the propagation direction $\hat y$. Since $\Delta y$ is small in this case, we take the principal value of the complex logarithm in Eq. (44).  For a homogeneous system, the extracted wave vector value would be independent of the location of the pair of detector points. We plot the extracted wave vector value for different sample points along the propagation directions. In this way, we can check how much the system diverges from the homogeneous case with increasing index contrast. For different pair of sampling points located at $y_{i}$, we plot the extracted real $k_{R}$ and imaginary part of $k_{I}$ of the wave vector k, for all five cases in Figs. 17(a) and 17(b) respectively. We find that for case (a) this value is almost constant. The small imaginary part present, is due to  numerical errors. Nonetheless, as the index contrast increases, deviations appear from a constant value  $k_{R}$, that become larger and larger. In addition, an increasingly large imaginary part appears, contradictory  to the fact that the photonic crystal is an inherently lossless system. Obviously, a phase velocity defined in the 2D crystal according to Yariv's picture \cite{yariv} quickly breaks down as the index contrast increases. Thus, for the large index contrast cases, the definition in Sec. IV for the phase velocity according to Notomi's picture \cite{Notomi} becomes appropriate.    
\par   
We now discuss the definition of  ``rightness'' for the low index cases. This property was found to have a sign that coincides with the curvature of the band for the cases where we have a strong mixing (high index contrast). As we have already mentioned, the band folding is just an artificiality when the medium is homogeneous or when the index contrast is very low. Negative curvature in such a case can by no means imply the existence of a left-handed (backwards) beam. The question arises when is it appropriate to associate the sign of band curvature with the sign of the ``rightness''? All beams in Figs. 13 (a) through (c) and  Figs. 16 (a) through (c) can very well be approximated by a plane wave. Therefore, clearly all the observed beams in such cases are right-handed beams. The cases of Figs.  13(d) and 16(d) have some small mixing, while the cases of Fig. 13(e)  and 16 (e) have a stronger mixing. As mentioned before, only in cases with a strong mixing does the phase velocity of the entire profile of the magnetic field  within the Wigner-Seitz cell have physical meaning. Apparently, only cases with a strong mixing, can be candidates for left-handed behavior. In fact, we should attempt to discuss ``rightness'' only for cases where the real part $k_{R}$ of the wave vector value (extracted using the methodology of the preceding paragraph) shows such large variations, that it ranges from positive to negative values. This is the case only for case (e) (index contrast 5.0:1.0). 
\par  
Conclusively, in the low index contrast cases, the wave vector diagram analysis is still valid. Even for a low index contrast in the higher bands, an effective medium cannot always describe the refractive properties of the photonic crystal. In an anisotropic homogeneous medium, at most, two beams would be expected but never three, as we observed in  Figs. 16 (b) through (e). However, it is inappropriate to assign a ``rightness'', according to the ``sign'' of the curvature of the band structures for PC's with low index modulation. Modulation must be high enough so that no predominant plane wave component exists. 
\par 
\vspace{1.5cm}
{\bf XI. COMPARISON WITH THE 1D LAYERED MEDIUM}
\par 
\vspace{0.2cm}
As we have mentioned before the refractive properties of the 1D layered medium have been extensively studied \cite{Russell1,Russell2,Russell3,Russell4,Zengerle}. However, there are significant differences between the properties of the 1D layered medium and the two-dimensional photonic crystal we studied in this paper. In the two-dimensional photonic crystal, when the plane of incidence is chosen to be the periodic plane, the entire wave vector is confined in the first BZ. In contrast, in the 1D periodic medium, only the component of the wave vector along the direction of periodicity is Bloch confined, i.e., restricted within the first BZ, in this case. This has several implications.
\par 
First, the modes in {\bf k}-space repeat themselves periodically in one direction only. If the interface is chosen along the 1D periodic direction, modes from the higher order zones, when $|{\bf k_{\|}} a/\pi|<1$, can never be accessed. One can access higher order modes for small $|{\bf k_{\|}}|$ values only when a slanted interface is employed. This is the method used in Refs. \cite{Russell1} and \cite{Russell2} to access modes lying outside the first BZ with small $|{\bf k_{\|}}|$. Note, in our 2D system we can access higher order modes even when the interface is cut along a symmetry direction and  small $|{\bf k_{\|}}|$. In fact, these are propagating waves indicated with dotted line in the cases of Fig. 1(b) and 1(c) (type I higher order waves).
\par
Second, the most important implication  is regarding the ``rightness.'' In the 2D system, a band with negative curvature corresponds to a left-handed (backwards) beam. However, this is not true for a one-dimensional system. We have chosen x to represent the direction of the periodicity. The curvature of a certain band will then be given by $\partial \omega(k_{x},k_{y})/\partial k_{x}$ or equivalently $v_{gx} k_{x}$, where $v_{gx}$ is the component of the group velocity along the direction of periodicity. Following the methods used in App. II for the group velocity and Sec. VII for the Poynting vector for the 1D layered system, we obtain,
\begin{eqnarray}
<{\bf S}> \cdot {\bf k}=\frac{c^2}{8\pi \omega} (k_{y}^2 A_{\eta}+\frac{\omega}{c^2} v_{gx} k_{x}),
\end{eqnarray}
where $<>$ refers to the spatial average within the unit cell of the time averaged quantity,  $\omega=\omega(k_{x},k_{y})$ for the band under consideration, and if the fields are H-polarized 
\begin{eqnarray}
A_{\eta}=\sum_{G,G'}{\eta_{G,G'} H_{G} H_{G'}}=\int {\eta(x) v^2 dx}.
\end{eqnarray}
$\eta(x)=\frac{1}{\epsilon(x)}$ and $v=\sum_{G}  {H_{G}(k_{x},k_{y}) \hspace{1mm} e^{iGx}}$ for the band under consideration. Since the integrand in expression (46) is positive, $A_{\eta}$ is a positive definite quantity \cite{Epol}.
\par
Now, for a band with a positive curvature, $v_{gx} k_{x}>0$ and so $<{\bf S}> \cdot {\bf k}>0$. For a band with a negative curvature $v_{gx} k_{x}<0$. Then $<{\bf S}> \cdot {\bf k}<0$, only if 
\begin{eqnarray}
k_{y}^2 A_{\eta}< \frac{\omega}{c^2} |v_{gx} k_{x}|,
\end{eqnarray}
where  $k_{x}$ is within the limits of the 1D BZ.
\par
In other words, a propagating wave that corresponds to a band with positive curvature is always a forward wave. In contrast, a propagating wave that corresponds to a band with a negative curvature is backwards (left-handed) only when condition (47) is satisfied. Note that condition (47) holds, regardless of the choice of the interface, provided that x represents the stacking (periodic) direction and y the direction perpendicular to this. If we choose the interface along y and consider normal incidence, then $k_{y}=0$. Thus, in this particular case, a band with negative curvature yields a backwards wave. Furthermore, we examined this condition for a case with high index modulation, H-polarization, frequency falling in the second band, and interface along the x-direction. With H-polarization in this case, we mean that the magnetic field is perpendicular to the plane of incidence. We employed the PWE method \cite{pwe1,villen,pwe2} and  found that the possibility of left-handed behavior, is restricted only to a small fraction of frequencies of the second band. In addition, for the applicable frequencies, one obtains backwards waves only for a part of the wave vector space. So, for the 1D layered medium, negative curvature does not necessarily imply a backwards beam. Each individual case should be examined with condition (47), to determine the ``rightness'' of the propagating beam. We note at this point that backwards coupling \cite{yariv3} observed between two waveguides linked with 1D layered medium does not necessarily imply a backwards wave.
\par
\vspace{1.5cm}
{\bf XII. CONCLUSIONS}
\vspace{0.2cm}
\par 
We systematically studied EM wave propagation in two-dimensional photonic crystal structures. We based our analysis on the wave vector diagram formalism. We observed different cases where negative refracted beam with distinctly different origins are present. We confirmed the condition for single beam propagation does not coincide with the condition for having a single reflected beam in the incoming medium. For simple cases, we determined the conditions for single beam propagation and applicability of Snell's formula. We revisited the controversial topic of phase velocity and showed that in a photonic crystal with strong scattering present,  only the wave vector inside the first BZ zone has physical meaning. We used the symmetrical properties of the photonic crystal to appropriately design a wedge experiment that can determine the ``rightness'' of a general 2D PC system (hexagonal or square). We studied the behavior of the PC system as the index contrast transitions from high to low values. We used, whenever possible, the symmetry of the system to determine the reflected beams by inferring a simple formula. For the cases in Sec. III the refracted beams were determined by the use of one primary construction line. With the rapid development of photonic crystals more complicated structures are now fabricated, for example 12-fold symmetrical quasi-crystals \cite{baumberg}. In more complicated structures, or when interfaces are not cut along crystal symmetry directions, the wave vector diagram analysis should be performed in its general form. All construction lines should be kept and all intersections should be accounted for to determine all possible reflected beams, as well as all  possible refracted beams. Our study will help in the understanding of EM propagation in more complicated and/or three dimensional structures. In 3D structures, interesting phenomena may arise because of the possibility of polarization coupling. The present study will also help in the understanding and  making of optical devices such as light deflection devices \cite{baba}, waveguide division multiplexers \cite {krauss} etc. 
\par
\vspace{1.5cm}
{\bf ACKNOWLEDGMENTS}
\par
\vspace{0.2cm}
We thank P. St. Russell and E. N. Economou for useful discussions. We also thank Kurt Busch for his help with the ${\bf k}\cdot {\bf p}$ perturbation method and useful discussion. Ames Laboratory is operated by the U.S Department of Energy by Iowa State University under Contract No. W-7405-Eng-82. This work was partially supported by DARPA (contract no. MDA972-01-2-0016) and a NSF international grant.
\par
\vspace{1.7cm}
{\bf APPENDIX I}\\ 
\par
\vspace{0.4cm}
{\bf a) Higher order Bragg reflected beams}
\vspace{0.2cm}
\par
The $k_{\|,m}$ component of an order m Bragg reflected wave in the air medium is given by Eq. (1). In order not to have any Bragg waves for any angle of incidence the condition,
\begin{eqnarray}
 |{\frac{\omega}{c} \sin \theta_{inc} + \frac{2 m \pi}{b_{str,cut}}}|> \omega/c,
\end{eqnarray}
must be observed $\forall \hspace{1mm} m \ne 0,\forall  \hspace{1mm} \theta_{inc}  \hspace{1mm} \epsilon  \hspace{1mm} [ 0,\pi]$. 
\par
But, if \begin{eqnarray}
\frac{\omega}{c} \sin \theta_{inc}> \frac{2 \pi}{b_{str,cut}},
\end{eqnarray}
then   

\begin{eqnarray}
|\frac{\omega}{c} \sin \theta_{inc}- \frac{2  \pi}{b_{str,cut}}}|= {\frac{\omega}{c} \sin \theta_{inc}- \frac{2  \pi}{b_{str,cut}} < \frac{\omega}{c} \sin \theta_{inc}<\frac{\omega}{c} ,
\end{eqnarray}
i.e., a Bragg wave of order m=-1 couples. 
Therefore, the first condition we must impose is

\begin{eqnarray}
 \frac{\omega}{c} \sin \theta_{inc} < \frac{2\pi}{b_{str,cut}}, 
\end{eqnarray} 
$ \forall  \hspace{1mm} |\theta_{inc}|<\pi$. Equivalently, 
\begin{eqnarray}
\frac{\omega}{c}< \frac{2\pi}{b_{str,cut}}
\end{eqnarray} 
must be satisfied. Assuming this condition is valid, we proceed.

\begin{eqnarray}
 |\frac{\omega}{c} \sin \theta_{inc}+ \frac{2 m \pi}{b_{str,cut}}| \geq min(|\frac{\omega}{c} \sin \theta_{inc}+ \frac{2 m \pi}{b_{str,cut}}|) \geq |\frac{\omega}{c}- \frac{2  \pi}{b_{str,cut}}|. 
\end{eqnarray}  
So, it suffices to find the frequencies to satisfy,
\begin{eqnarray}
|\frac{\omega}{c}- \frac{2  \pi}{b_{str,cut}}|>\frac{\omega}{c}
\end{eqnarray}
or equivalently,
\begin{eqnarray} 
\tilde f \leqslant \frac{a}{2 b_{str,cut}},
\end{eqnarray}
with a being the lattice constant. 
\par
 No higher order reflected beams appear for any angle of incidence for frequencies satisfying (55). If condition (55) (or (54)) is valid, condition (52) is automatically satisfied. In addition, for a certain frequency satisfying (52), for a  certain incident angle $\theta_{inc}$ obeying the inequality,
\begin{eqnarray} 
-\frac{\omega}{c} \sin \theta_{inc} + {2  \pi}/{b_{str,cut}}>\frac{\omega}{c},
\end{eqnarray}
no higher order Bragg reflected beams appear.
\vspace{0.4cm}

{\bf b) Equivalent points in wave vector space} \\
\vspace{0.2cm}
The eigenvalue equation that results from expression (4) and Maxwell's equation is for the H polarization case,
\begin{eqnarray}
\sum_{\bf G^{'}}{\eta_{\bf G,G'} ({\bf k}+{\bf G}) \cdot  ({\bf k}+{\bf G^{'}}) H_{\bf G^{'}} ({\bf k})}=\frac{\omega^{2}}{c^2} H_{{\bf G}} ({\bf k}).
\end{eqnarray}
\par
 Suppose we consider the FB wave for ${\bf K}={\bf k}+{\bf G_{o}}$ with ${\bf G_{o}}$ a reciprocal lattice vector. Then, the eigenvalue equation becomes
\begin{eqnarray}
\sum_{\bf G^{'}}{\eta_{\bf G,G'} ({\bf k}+{\bf G}+{\bf G_{o}}) \cdot  ({\bf k}+{\bf G^{'}}+{\bf G_{o}}) H_{{\bf G^{'}}}({\bf K})}=\frac{\omega^{2}}{c^2} H_{{\bf G}}({\bf K}).
\end{eqnarray}
 From Eq. (58) after setting ${\bf G_{1}}={\bf G}+{\bf G_{o}}$ and ${\bf G_{2}}={\bf G^{'}}+{\bf G_{o}}$ we get 
\begin{eqnarray}
\sum_{\bf G_{2}}{\eta_{\bf G_{1},G_{2}} ({\bf k}+{\bf G_{1}}) \cdot  ({\bf k}+{\bf G_{2}}) H_{{\bf G_{2}}-{\bf G_{0}}} ({\bf K})}=\frac{\omega^{2}}{c^2} H_{{\bf G_{1}}-{\bf G_{0}}} ({\bf K}). 
\end{eqnarray}
By comparison with the original eigenvalue equation, it is evident that
\begin{eqnarray}
H_{{\bf G}-{\bf G_{0}}} ({\bf K})=H_{{\bf G}} ({\bf k}).
\end{eqnarray}
\par 
Therefore, the time-independent part of the FB wave (Eq. (4)) for {\bf K} becomes 
\begin{eqnarray}
H_{FB,{\bf K}}=e^{i {\bf K} \cdot {\bf  r}} \sum_{\bf G} {H_{{\bf G}} ({\bf K})  \hspace{1mm} e^{i {\bf G} \cdot {\bf r}} }
= e^{i {{\bf k}\cdot{\bf  r}}} \sum_{\bf G}{H_{{\bf G}} ({\bf K}) \hspace{1mm} e^{i({\bf G}+{\bf G_{o}}) \cdot {\bf r}}}\nonumber
\end{eqnarray}
\begin{eqnarray}
=e^{i{\bf k}\cdot {\bf  r}} \sum_{\bf G^{'}} {H_{{\bf G^{'}}-{\bf G_{o}}} ({\bf K})  \hspace{1mm} e^{i{\bf G^{'}}\cdot{\bf r}}}
=e^{i{\bf k}\cdot {\bf r}} \sum_{\bf G^{'}} {H_{{\bf G^{'}}} ({\bf k}) \hspace{1mm} e^{i {\bf G^{'}} \cdot {\bf r}}} = H_{FB,{\bf k}}.
\end{eqnarray}
So, the Floquet-Bloch wave expressions corresponding to wave vectors {\bf k} and {\bf K}, separated by a reciprocal lattice vector ${\bf G_{o}}$, are equivalent. In other words, {\bf k} and {\bf K} are equivalent points in wave vector space.
\par 
\vspace{0.4cm}
{\bf c) Yariv's definition for phase velocity}
\par
\vspace{0.2cm}
In the 1D system the wave vector in the plane of incidence is not confined in the first BZ, but only the component along the periodicity. Therefore, the FB wave has the form (when magnetic field perpendicular to the plane of incidence)
\begin{eqnarray} 
{\bf H}({\bf r},t)= e^{i kx} e^{i \beta y} \sum_{G}{H_{{\bf G}} ({ k},\omega) \hspace{1mm} e^{i G x}} \hspace{0.5mm} e^{-i\omega t} \hspace{1mm} \hat z.
\end{eqnarray} 
 x is chosen to represent the direction of periodicity.  The phase velocity defined in Ref. \cite{yariv} for the FB wave, given by (62) is
\begin{eqnarray}
v_{p}=\frac{c}{\sqrt{k^2+\beta^2}}. 
\end{eqnarray}
In this expression (see Ref.\cite{yariv}), k is not within the first BZ zone, but chosen so that $|H_{0}|>$$|H_{G}|$ $\forall G \hspace{1mm} \ne 0$. 
\vspace{0.4cm}
\par
{\bf d) The average of the FB wave in the Wigner-Seitz cell}
\par
\vspace{0.2cm} 
\begin{eqnarray}
<H( {\bf r}={\bf R})>=\frac{1}{\sqrt {A_{WS}}} \hspace{0.5mm} e^{-i\omega t} \newline \int_{{\bf r'}} \hspace{0.5mm} { e^{i{\bf k} \cdot ({\bf R}+{\bf r'})} 
\sum_{\bf G}{H_{{\bf G}}e^{i{\bf G} \cdot ({\bf R}+{\bf r'})}} d^{2}{\bf r'}} \nonumber
\end{eqnarray}
\begin{eqnarray}
=\frac{1}{\sqrt {A_{WS}}} \hspace{0.5mm} e^{-i\omega t} e^{i{\bf k} \cdot {\bf R}} \newline \int_{{\bf r'}} \hspace{0.5mm} { e^{i{\bf k} \cdot {\bf r'}} 
\sum_{\bf G}{H_{{\bf G}} e^{i{\bf G} \cdot {\bf r'}}} d^{2}{\bf r'}}
=\hspace{1mm} e^{-i\omega t} e^{i{\bf k} \cdot {\bf R}}<H( {\bf r}={\bf 0})>
\end{eqnarray}
\newline
where ${\bf r^{'}}$ ranges within the Wigner-Seitz cell around {\bf r}={\bf R}, {\bf R} a Bravais lattice vector, and we used $e^ {i {\bf G}\cdot {\bf R}}$=1.

\par 
\vspace{1.5cm}
{\bf Appendix II: Calculation of the group velocity}\\
\par
\vspace{0.2cm}
For the group velocity calculation we employ the {$\bf k \cdot \bf p$} perturbation method that was first introduced for the photonic crystal by Johnson et al. \cite{kp_phot}. Busch et al. \cite{Busch1,Busch2} implemented this method to determine the group velocity and ``photon mass'' of a two-dimensional E-polarized photonic crystal. In the following, we provide the basic steps of such calculation and derive simple expressions in terms of the FB wave's coefficients. The final expressions are compared with the corresponding energy velocity expressions in Sec. VII.\\
\vspace{1cm}
{\bf a) H-polarization} \\
From Maxwell's equations we obtain:\\
\begin{eqnarray}
\nabla \times (\frac{1}{\epsilon({\bf r})} \nabla \times {\bf H}_{n, {\bf k}})=\frac{\omega_{n,{\bf k}}^{2}}{c^2} {\bf H}_{n, {\bf k}},
\end{eqnarray}
where ${\bf H}_{n, {\bf k}}$ is the FB wave that corresponds to a band with index n, wave vector within the first BZ {\bf k} and frequency $\omega_{n{\bf k}}$. The FB wave is given by expression (17) with $v_{n,\bf k}$ given by expression (19). Since the wave is H-polarized, the eigenvectors are parallel to the cylinder axis z, and {\bf k} lies in the plane of periodicity xy. Using expression (65), (17) and (19) we get\\
\begin{eqnarray}
\nabla \times (\frac{1}{\epsilon({\bf r})} \nabla \times (e^{i {\bf k \cdot r}} {\bf v}_{n, {\bf k}}))=\frac{\omega_{n, {\bf k}}^{2}}{c^2} \hspace{1mm} e^{i {\bf k \cdot r}} {\bf v}_{n,{\bf k}}.
\end{eqnarray}  
We set ($\eta({\bf r})=\frac{1}{\epsilon({\bf r})}$), and take into account that ${\bf k} \cdot  {\bf v}_{n, {\bf k}}=0$ and that ${\bf v}_{n,{\bf k}}={\bf v}_{n,{\bf k}} (x,y)$ (since xy is the plane of periodicity). After manipulations, we get\\

\begin{eqnarray}
{\bf \hat O_{k}} {\bf v}_{n, {\bf k}}=\lambda_{n, {\bf k}} {\bf v}_{n,{\bf k}},
\end{eqnarray}
with\\

\begin{eqnarray}
{\bf \hat O_{k}}=\nabla \eta({\bf r}) \times \nabla \times + i \nabla \eta({\bf r}) \times {\bf k} \times+ 
\eta({\bf r}) [ i {\bf k} \times \nabla \times +k^{2}-\nabla^{2}-i({\bf k} \cdot \nabla)]
\end{eqnarray}
and
\begin{eqnarray}
\lambda_{n,{\bf k}}=\frac{\omega_{n,{\bf k}}^{2}}{c^2}.
\end{eqnarray}

Equation (67) represents the eigenvalue equation that yields the FB wave states defined in Eq. (4) for a certain wave vector ${\bf k}$ and frequency $\omega_{n, {\bf k}}$. Let us now consider the FB wave for the wave vector ${\bf k^{'}={\bf k}+{\bf q}}$ where $|{\bf q}|<< \pi/a$ . The eigenvalue equation for wave vector value ${\bf k}+{\bf q}$ will be\\
\begin{eqnarray}
{\bf \hat O}_{{\bf k}+{\bf q}} {\bf v}_{n, {\bf k}+{\bf q}}=\lambda_{n,{\bf k}+{\bf q}}{\bf v}_{n, {\bf k}+{\bf q}}.
\end{eqnarray}
Setting ${\bf k} \rightarrow {\bf k}+{\bf q}$ in Eq. (68) we get:\\
\begin{eqnarray}
{\bf \hat O_{{\bf k}+{\bf q}}}={\bf \hat O}_{{\bf k}}+ {\bf q}\cdot {\bf \Omega}+O(\epsilon^{2})
\end{eqnarray}
with
\begin{eqnarray}
{\bf \Omega}=-i \nabla \eta({\bf r})-2 i \eta({\bf r}) \nabla+2 \eta({\bf r}) {\bf k}.
\end{eqnarray}
So, the eigenvalue operator for ${\bf k}+{\bf q}$ can be considered as the corresponding operator for  ${\bf k}$ with a perturbation ${\bf q}\cdot {\bf \Omega}$. Thus,
\begin{eqnarray}
\lambda_{n,{\bf k}+{\bf q}}=\lambda_{n,{\bf k}} +{\bf q} \cdot \frac{<{\bf v}_{n, {\bf k}}| {\bf \Omega}|{\bf v}_{n, {\bf k}}>}{<{\bf v}_{n, {\bf k}}|{\bf v}_{n, {\bf k}}>}
\end{eqnarray} 
Consequently,
\begin{eqnarray}
\omega_{n,{\bf k}+{\bf q}}-\omega_{n,{\bf k}}={\bf q} \hspace{2mm} \cdot \nabla_{{\bf k}} \omega= 
\frac{c^{2}}{2 \omega_{n,{\bf k}}} {\bf q} \cdot \frac{<{\bf v}_{n, {\bf k}}|\hspace{1mm} {\bf \Omega} \hspace{1mm}|{\bf v}_{n, {\bf k}}>}{<{\bf v}_{n, {\bf k}}|{\bf v}_{n, {\bf k}}>}.
\end{eqnarray}
Apparently,\\
\begin{eqnarray}
{\bf v_{g}}=\frac{c^{2}}{2 \omega_{n,{\bf k}}} \frac{<{\bf v}_{n, {\bf k}}|\hspace{1mm}{\bf \Omega}\hspace{1mm}|{\bf v}_{n, {\bf k}}>}{<{\bf v}_{n, {\bf k}}|{\bf v}_{n, {\bf k}}>}.
\end{eqnarray}
Note that $<{\bf v}_{n, {\bf k}}|\hspace{1mm}{\bf \hat O}\hspace{1mm}|{\bf v}_{n, {\bf k}}>=\int_{WSC}\hspace{1mm} {{\bf v}_{n, {\bf k}}^{*} \hspace{1mm} \hat O \hspace{1mm} {\bf v}_{n, {\bf k}}} \hspace{1mm} d^2{\bf r}$. The integral is over the two-dimensional Wigner-seitz (WS) cell. Each term of the operator ${\bf \Omega}$ is evaluated separately in the numerator of the expression (75). In fact, using expression (19) for ${\bf v}_{n, {\bf k}}$ we obtain
\begin{eqnarray}
<{\bf v}_{n, {\bf k}}|-i \nabla \eta({\bf r})|{\bf v}_{n, {\bf k}}>=\sum_{{\bf G},{\bf G^{'}}} {{\bf G^{'}} \eta_{{\bf G^{'}}} H_{{\bf G}+{\bf G^{'}}} H_{{\bf G}}}
\end{eqnarray}
\begin{eqnarray}
<{\bf v}_{n, {\bf k}}|-2 i \eta({\bf r}) \nabla|{\bf v}_{n, {\bf k}}>=2  \sum_{{\bf G},{\bf G^{'}}} {{\bf G} \hspace{1mm} \eta_{{\bf G^{'}}} H_{{\bf G}+{\bf G^{'}}} H_{{\bf G}}}
\end{eqnarray}
\begin{eqnarray}
<{\bf v}_{n, {\bf k}}|2 \eta({\bf r}) {\bf k}|{\bf v}_{n, {\bf k}}>=2  \sum_{{\bf G},{\bf G^{'}}} {{\bf k} \hspace{1mm} \eta_{{\bf G^{'}}} H_{{\bf G}+{\bf G^{'}}} H_{{\bf G}}}
\end{eqnarray} 
and
\begin{eqnarray}
<{\bf v}_{n, {\bf k}}|{\bf v}_{n, {\bf k}}>=\sum_{{\bf G}}{ H_{{\bf G}}^{2}}=1.
\end{eqnarray} 
(eigenvectors are normalized to unity). Note to derive the above relations we used
\begin{eqnarray}
\frac{1}{A_{WSC}} \int_{WSC} { e^{i {\bf G} \bf r}} d^{2} {\bf r}=\delta({\bf G}).
\end{eqnarray}
Finally, lumping all terms together we calculate\\

\begin{eqnarray}
{\bf v_{g}}=\frac{c^2}{2 \omega_{n,{\bf k}}} \sum_{{\bf G},{\bf G^{'}}} { (2 {\bf k}+2 {\bf G}+{\bf G^{'}}) \hspace{1mm} \eta_{{\bf G^{'}}} H_{{\bf G}} H_{{\bf G}+{\bf G^{'}}}}.
\end{eqnarray}
After index manipulation (i.e., setting ${\bf G_{1}}={\bf G}+{\bf G^{'}}$ and ${\bf G_{2}}={\bf G}$), we obtain
\begin{eqnarray}
{\bf v_{g}}=\frac{c^2}{\omega_{n,{\bf k}}} \sum_{{\bf G_{1}},{\bf G_{2}}} { ( {\bf k}+ {\bf G_{1}}) \hspace{1mm} \eta_{{\bf G_{1}}-{\bf G_{2}}} H_{{\bf G_{1}}} H_{{\bf G_{2}}}}.
\end{eqnarray}
To obtain the above expression, we used
\begin{eqnarray}
\sum_{{\bf G_{1}},{\bf G_{2}}} { \hspace{2mm} {\bf G_{1}} \hspace{2mm} \eta_{{\bf G_{1}}-{\bf G_{2}}} H_{{\bf G_{1}}} H_{{\bf G_{2}}}}= 
\sum_{{\bf G_{1}},{\bf G_{2}}} \hspace{2mm}{ {\bf G_{2}} \hspace{2mm} \eta_{{\bf G_{1}}-{\bf G_{2}}} H_{{\bf G_{1}}} H_{{\bf G_{2}}}},
\end{eqnarray}
since the expression inside the sum is symmetrical for the pair of reciprocal vectors ${\bf G_{1}},{\bf G_{2}}$. 
\par
\vspace{0.4cm} 
{\bf b) E-polarization} 
\vspace{0.2cm}
\par 
From Maxwell's equations,
\begin{eqnarray}
\nabla \times \nabla \times {\bf E}=\frac{\omega_{n,{\bf k}}^2}{c^2} \epsilon({\bf r}) {\bf E}.
\end{eqnarray}
The electric field satisfies Bloch's theorem, therefore,\\
\begin{eqnarray}
 {\bf E}=e^{i {\bf k}\cdot {\bf r}} {\bf u}_{n,{\bf k}},
\end{eqnarray}
with 
\begin{eqnarray} 
{\bf u}_{n,{\bf k}}=\frac{1}{\sqrt{A_{WSC}}} \sum_{{\bf G}} { {\bf E}_{\bf G} e^{i {\bf G}\cdot {\bf r}}}.
\end{eqnarray}

Note in this case the electric field {\bf E} is parallel to the cylinder axis z. This means ${\bf k} \cdot {\bf u}_{n,{\bf k}}=0$ and ${\bf E}_{\bf G}=E_{\bf G} \hspace{1mm} \hat z$. Expression (84) with (85) and (86) yields
\begin{eqnarray}
{\bf \hat O_{k}} {\bf u}_{n,{\bf k}}=\lambda_{n, {\bf k}} {\bf u}_{n,{\bf k}},
\end{eqnarray}
with \cite{Busch1,Busch2}
\begin{eqnarray}
{\bf \hat O_{k}}=-\nabla^{2}-2 i({\bf k \cdot \nabla})+ k^{2},
\end{eqnarray}
and $\lambda_{n,{\bf k}}$ given by (69).
After following a similar analysis as in a), we obtain
\begin{eqnarray}
{\bf v_{g}}=\frac{c^2}{2 \omega_{n,{\bf k}}} \frac{<{\bf u}_{n,{\bf k}}|{\bf \Omega}|{\bf u}_{n,{\bf k}}>}{<{\bf u}_{n,{\bf k}}|\epsilon({\bf r})|{\bf u}_{n,{\bf k}}>}
\end{eqnarray}
with   
\begin{eqnarray}
{\bf \Omega}=-2i \nabla+2 {\bf k}.
\end{eqnarray}
Expression (89), after acting ${\bf \Omega}$ on ${\bf u}_{n,{\bf k}}$ yields
\begin{eqnarray}
{\bf v_{g}}=\frac{c^2}{\omega_{n,{\bf k}}} \sum_{{\bf G}} { ({\bf k}+{\bf G}) E_{{\bf G}}^2}. 
\end{eqnarray}

\begin{figure}[b] 
\begin{center} 
\includegraphics[angle=0,width=13cm,angle=270]{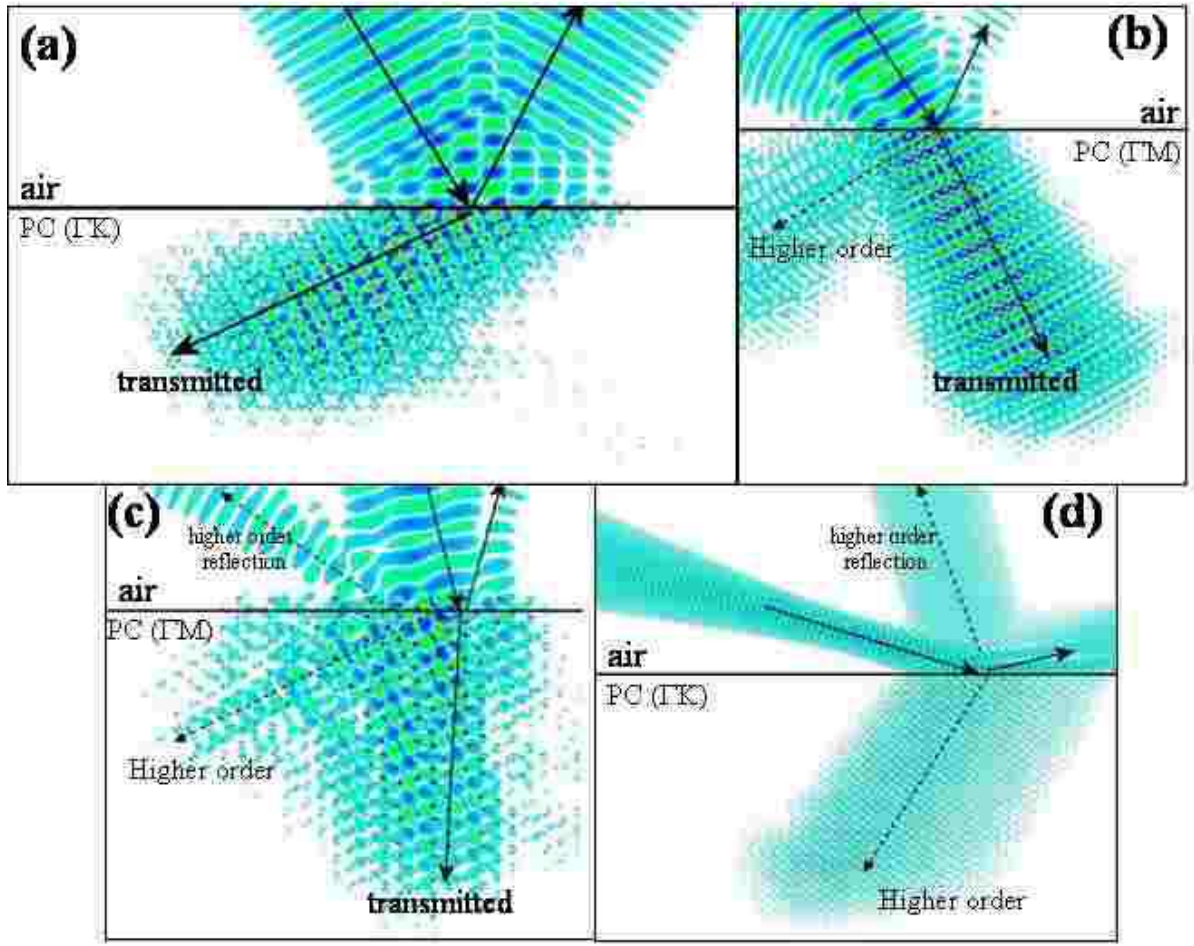}
\end{center}
\caption{Oblique incidence of EM waves at photonic crystal slabs. The PC system consists of dielectric rods in a hexagonal arrangement. All cases are with magnetic field along the cylinders (H-polarization). The parameters for each case (dielectric constant of rods $\epsilon$, radius of rods r, and dimensionless frequency $\tilde f$) are:  a) $\epsilon$=12.96,  $r=0.35 a$, $\tilde f$=0.58,  b) $\epsilon$=20.0, $r=0.37 a$, $\tilde f$=0.425, c) $\epsilon$=12.96, $r=0.35 a$, $\tilde f$=0.535, and finally in d) $\epsilon$=7.0, $r=0.35 a$ , $\tilde f$=0.81. Note that $\tilde f$=$fa/c=a/\lambda$, with $a$ the lattice constant and c the velocity of light and $\lambda$ the wavelength of light in vacuum. The solid arrows indicate the transmitted, while the dotted black arrows indicate higher order beams inside the PC.}
\end{figure} 
\begin{figure}[b] 
\begin{center}
\includegraphics[angle=0,width=14cm,angle=0]{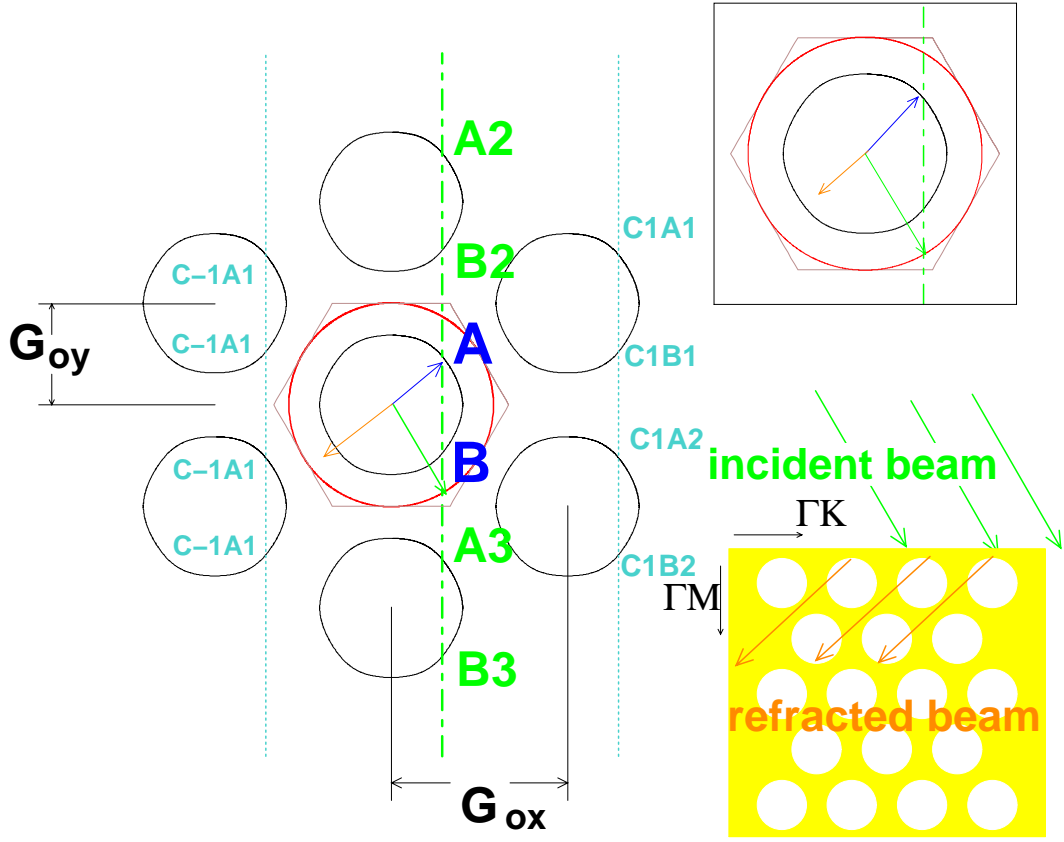}
\end{center}
\caption{Wave vector diagram for the case of Fig. 1(a). The equifrequency surfaces are plotted (black solid lines) in the repeated zone scheme. The red solid circle represents the equifrequency surface for the air (incoming) medium. The green dot-dashed line is the primary (zeroth order) construction line. The additional turquoise lines represent higher order construction lines. All intersections are indicated. The blue vector represents the fundamental wave vector of the FB wave that corresponds to a causal signal. The respective energy velocity that coincides with the propagating signals direction is shown as the orange vector. In the insert the corresponding wave vector analysis in the first zone is shown.  Clearly, an analysis in the first zone is sufficient in this case.  For this cut ${\bf G_{ox,y}}=2 \pi/a_{x,y}$ with $a_{x}=a$ (lattice constant) and $a_{y}=\sqrt 3 a$. A general reciprocal lattice vector is (2n1+n2) ${\bf G_{ox}}$ + n2 ${\bf G_{oy}}$, with n1, n2 integers.}
\end{figure}
\begin{figure}[b] 
\begin{center}
\includegraphics[angle=0,width=14cm]{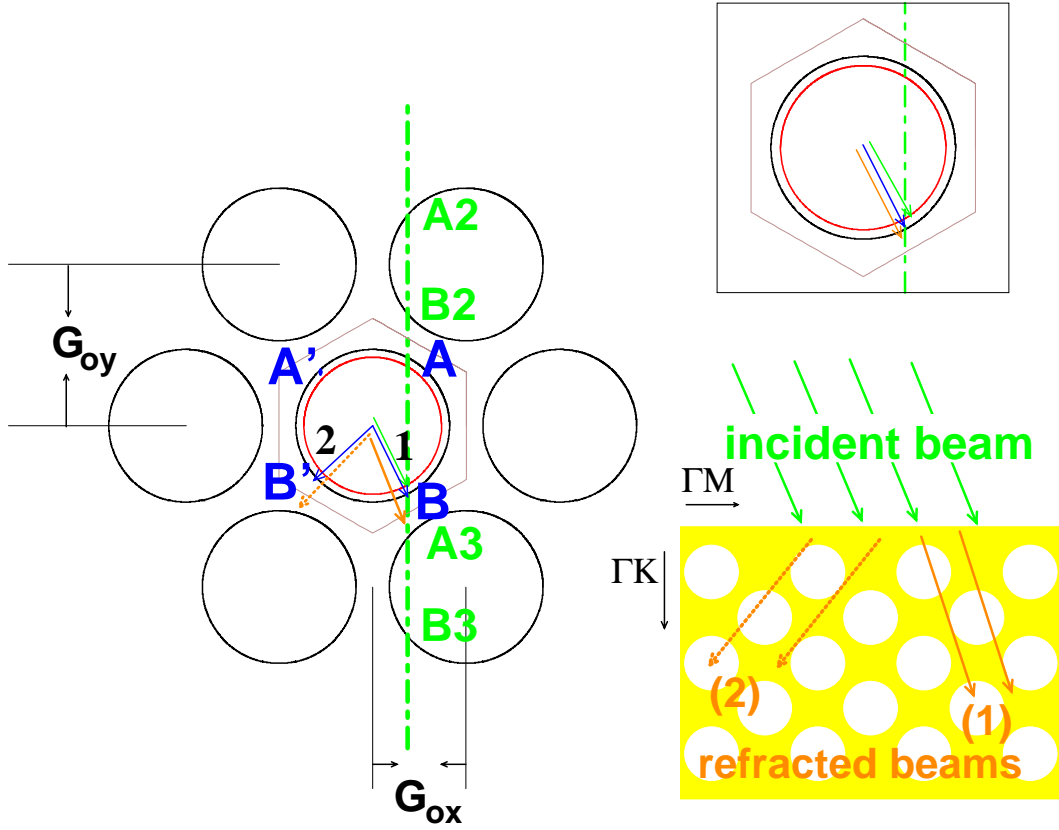}
\end{center}
\caption{Wave vector diagram for the case of Fig. 1(b). The equifrequency surfaces are plotted (black solid lines) in the repeated zone scheme. The red solid circle represents the equifrequency surface for the air (incoming) medium. The green dot-dashed line is the primary (zeroth order) construction line. All intersections are indicated. The blue vectors represent the intersections that result in the first zone after the folding process, that correspond to causal signal (shown as the orange vectors). In the insert, the corresponding wave vector analysis done in the first zone is shown,. Thus, the latter fails to predict the second refracted beam. For this cut ${\bf G_{ox,y}}=2 \pi/a_{x,y}$ with \hspace{1mm} $a_{x}=\sqrt 3 a$ (lattice constant) and $a_{y}=a$. A general reciprocal lattice vector is  n1 ${\bf G_{ox}}$ + (n1+2n2) ${\bf G_{oy}}$, with n1, n2 integers.}
\end{figure}
\begin{figure}[b]  
\begin{center}
\includegraphics[angle=0,width=14cm]{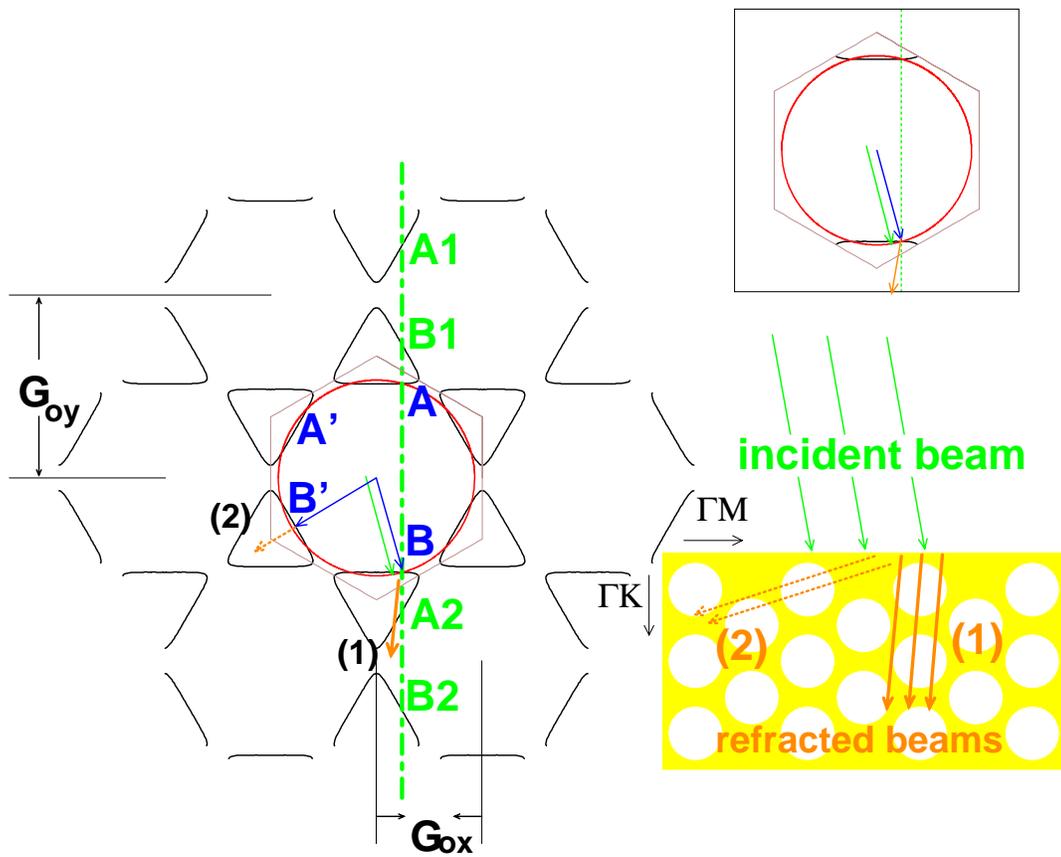}
\end{center}
\caption{Same as Fig. 3 but for the case of Fig. 1(c).} 
\end{figure} 
\begin{figure}[b] 
\begin{center} 
\includegraphics[angle=0,width=14cm]{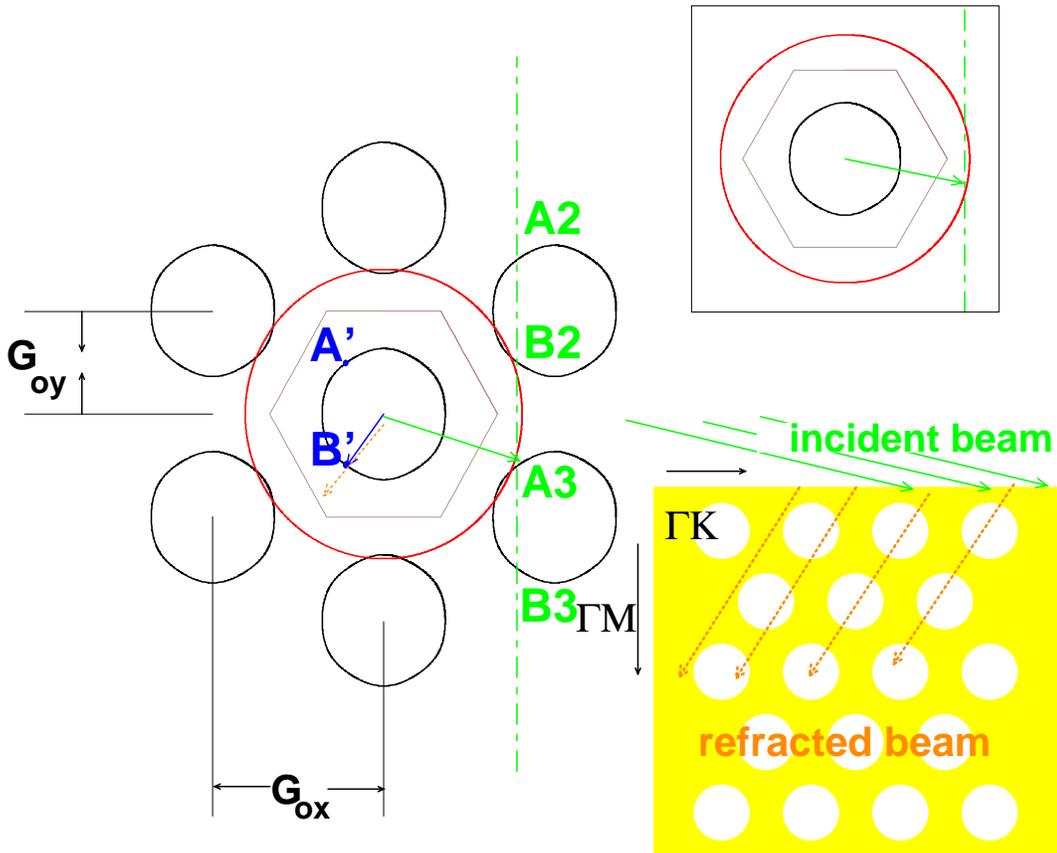}
\caption{Wave vector diagram for the case of Fig. 1(d) in the extended zone scheme. For this cut ${\bf G_{ox,y}}=2 \pi/a_{x,y}$ with $a_{x}=a$ (lattice constant) and $a_{y}=\sqrt 3 a$. A general reciprocal lattice vector is (2n1+n2) ${\bf G_{ox}}$ + n2 ${\bf G_{oy}}$, with n1, n2 integers. Everything else is the same as in the previous figures. Notice that in this case a wave vector type of analysis in the first zone (see insert) predicts no propagating signal.}
\end{center}
\end{figure}  
\begin{figure}[b] 
\begin{center} 
\includegraphics[width=10cm]{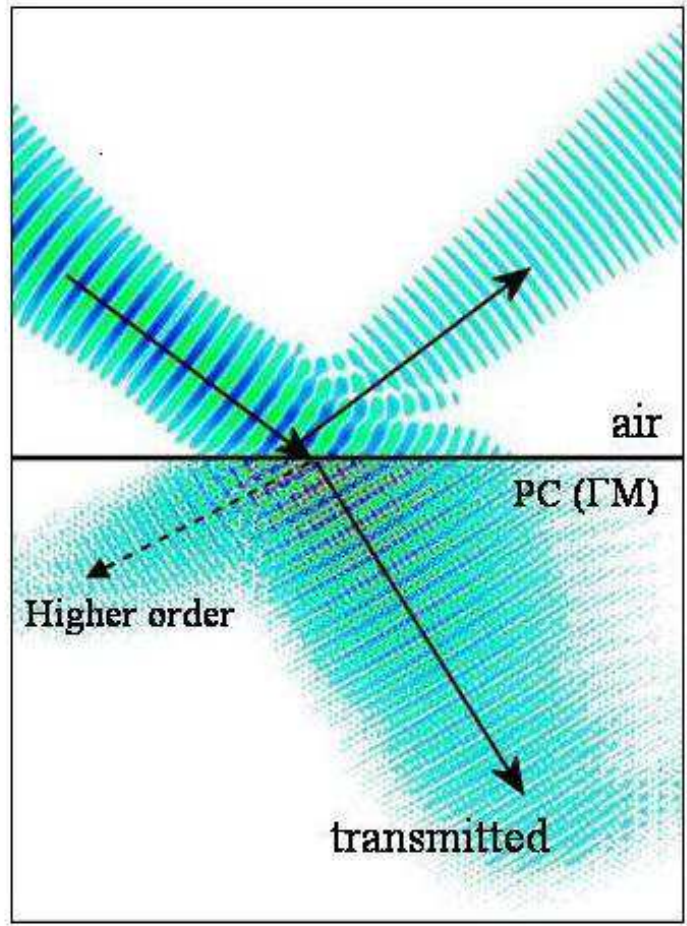}
\end{center}
\caption{Oblique incidence at the photonic crystal slab with $\epsilon=60.$, $r=0.37 a$, for frequency $\tilde f=0.275$ that is below the Bragg condition for no additional reflected beams for any angle of incidence. Notice that despite the presence of only one reflected beam, there are two propagating beams indicated with the black solid and dotted arrows, respectively. } 
\end{figure}
\begin{figure}[b] 
\begin{center}
\includegraphics[width=14cm,angle=0]{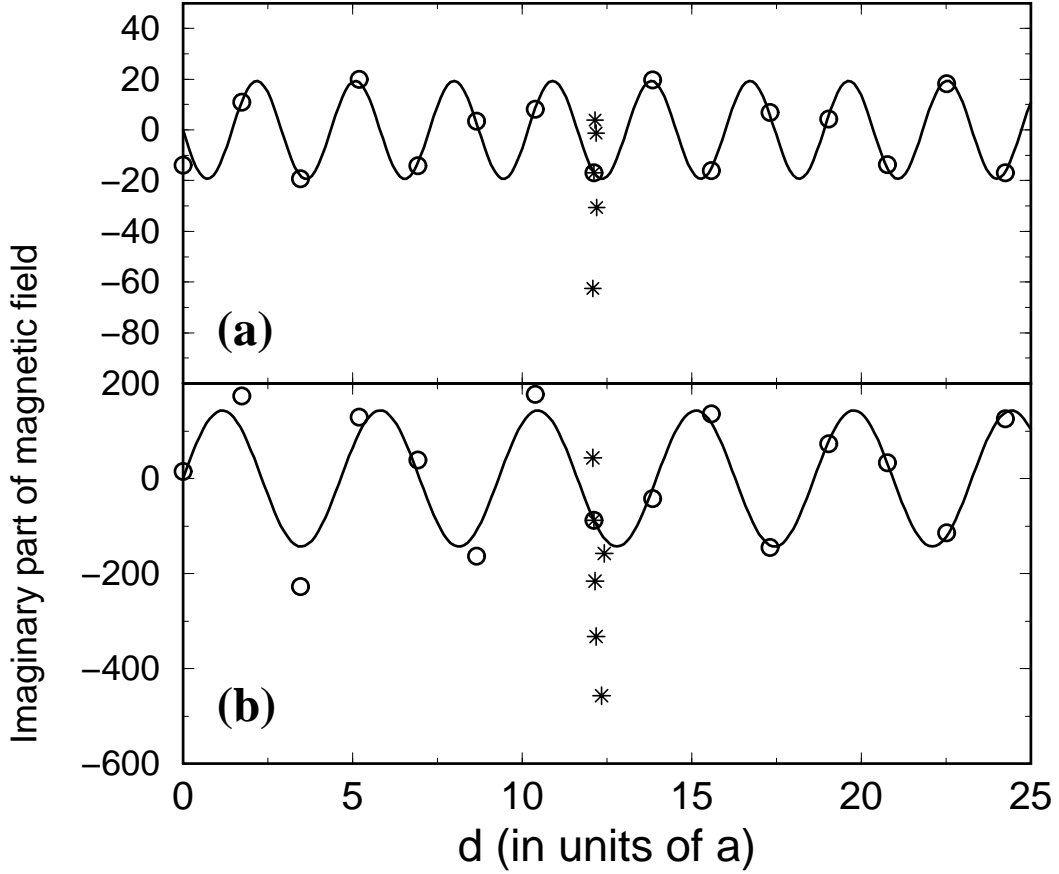}
\end{center}
\caption{The open circles represent the imaginary part of the Fourier transformed magnetic field, sampled in time at different points along the propagation direction $y_{i}$. The sampling points are separated by one period along the propagation direction which is the $\Gamma M$ direction. PC has rods with $\epsilon=12.96$ and radius 0.35 a, and the magnetic field lies along the cylinders (H-polarization). In both cases, the corresponding equifrequency surfaces are almost isotropic. Top panel (a) is for $\tilde f$=0.58 that belongs to a band with negative curvature. The bottom panel  is for $\tilde f$=0.48 that belongs to a band with positive curvature. The solid lines are $\propto \sin(k y_{i})$, where k is the real part of the numerically calculated wave vector. Thus, k=-2.16 $a^{-1}$ for case (a) and 1.35 $a^{-1}$ case (b). The stars represent the imaginary part of the Fourier transformed magnetic field for points in the neighborhood of $y_{i}=7 b$ with $b=\sqrt 3$. } 
\end{figure}
\begin{figure}[b]
\begin{center}
\includegraphics[width=14cm,angle=0]{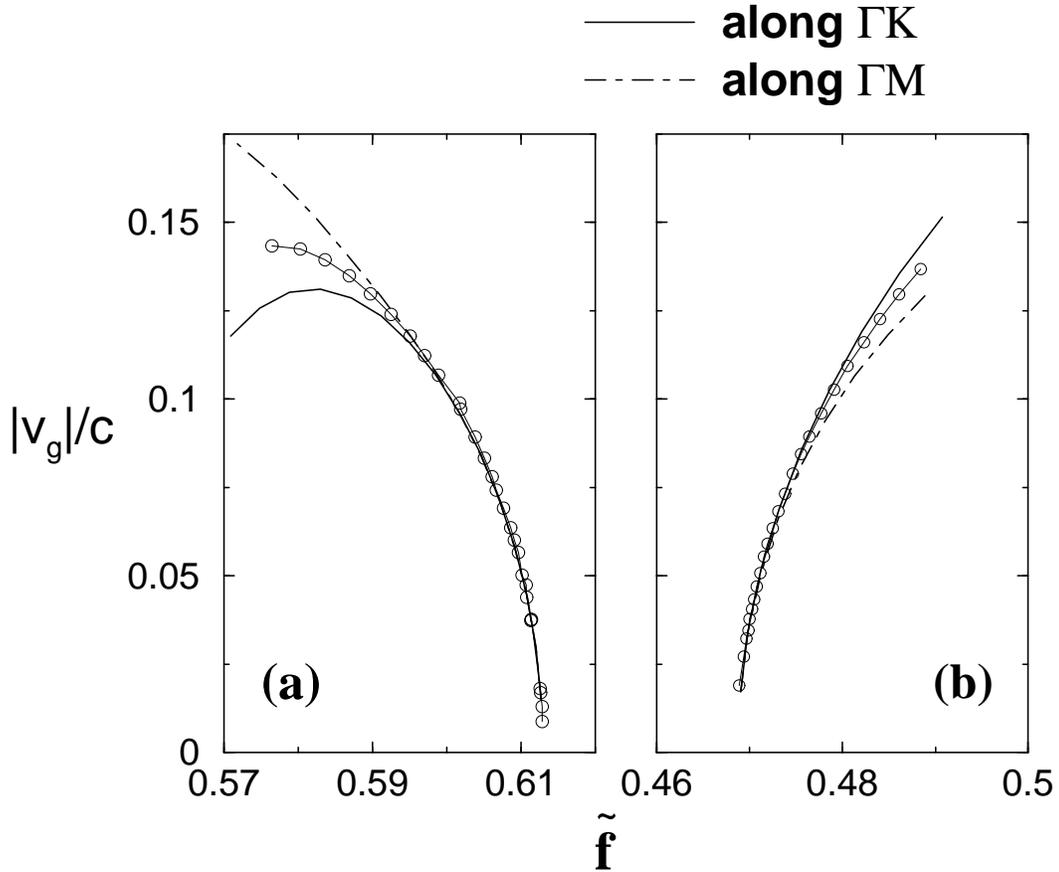}
\end{center}
\caption{Magnitude of the group velocity for cases with almost isotropic EFS for a photonic crystal of rods with $\epsilon=12.96$ and radius 0.35, for H-polarization. In a) the results of the $5^{th}$ band (negative curvature band) are shown. In b) the results of the $4^{th}$ band (positive curvature band) are shown. The solid and dot-dashed lines represent the results from the ${\bf k \cdot p}$ perturbation method for signal along the $\Gamma K$ and  $\Gamma M$ direction, respectively. The solid lines with circles represent the results obtained when considering the system having an effective phase index $n_{p}$. The index is calculated from the band structure (EFS surfaces) and is frequency dependent. Agreement between the two results is excellent close to the band edge. Since the anisotropy increases as one moves away from the band edge, so does the discrepancy between the two values.} 
\end{figure}
\begin{figure}[b] 
\begin{center}
\includegraphics[width=14cm,angle=0]{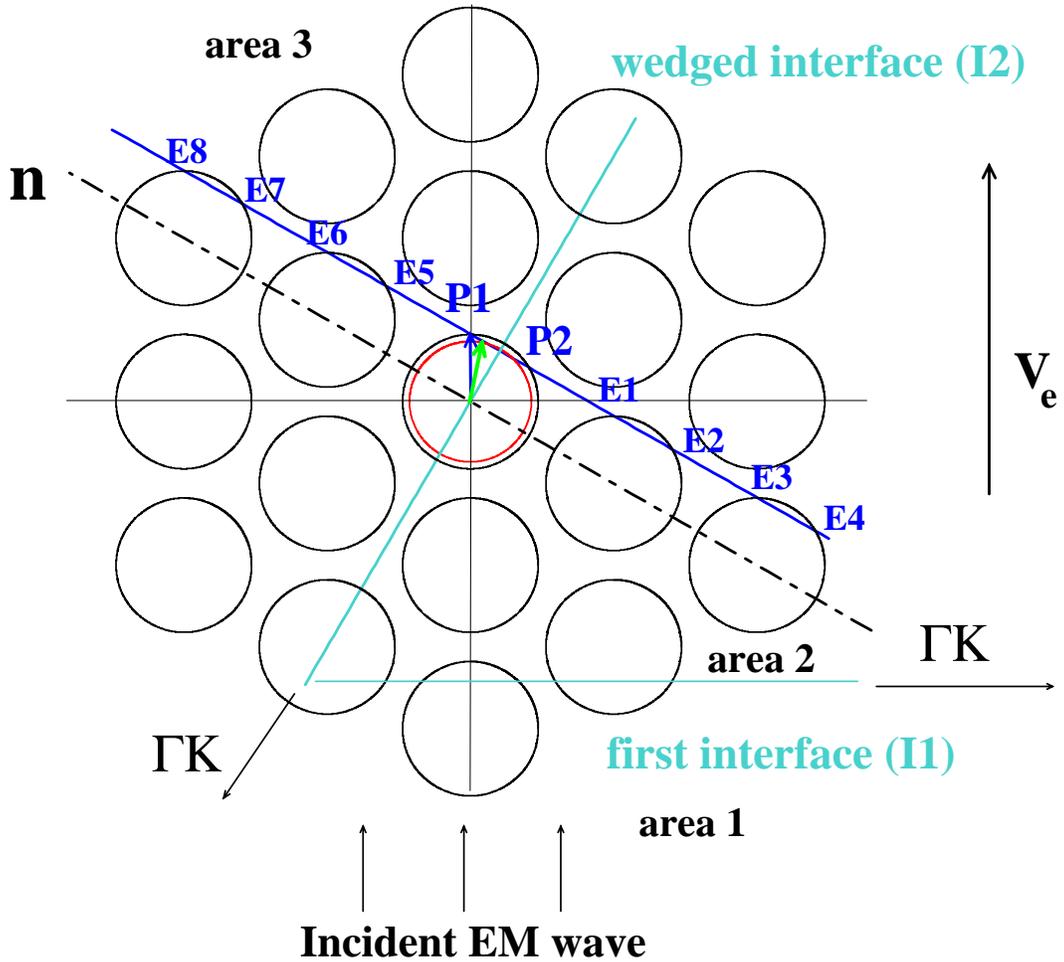}
\end{center}
\caption{Wave vector diagrams for the wedge experiment that corresponds to Fig. 1(b). The black solid lines represent the EFS in the PC medium. The red circle is the EFS in air. The blue vector represents the wave vector inside the PC medium. The black solid arrow represents the causal direction of the energy flow inside the PC. The turquoise lines indicate the directions of the first and wedged interfaces, both chosen along $\Gamma K$. The normal to the wedged interface is indicated with the dot-dashed line. The blue solid line indicates the construction line at the wedged interface. The green arrow represents the wave vector and direction of the outgoing beam (in air).}
\end{figure}
\begin{figure}[b] 
\begin{center}
\includegraphics[width=10cm]{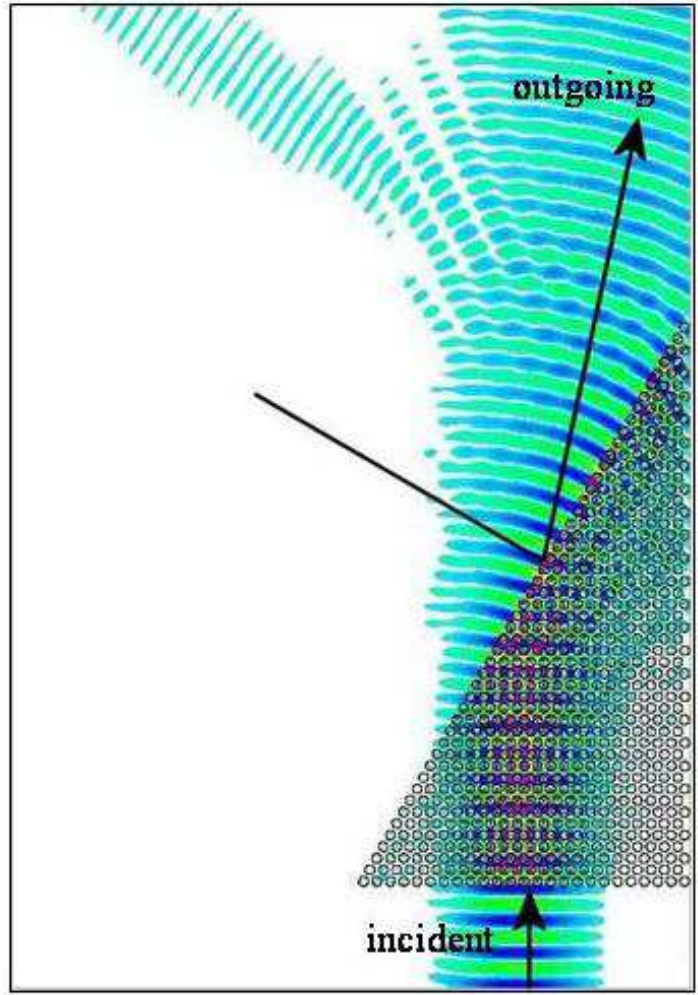}
\end{center} 
\caption{The wedge experiment for the case of Fig. 1(b). The operation frequency is $\tilde f=0.425$. A large positive outgoing angle is seen as predicted by the wave vector diagram analysis. The second beam is along the normal to the wedge direction and is due to multi-reflections in the upper part of the wedge.}
\end{figure}
\newpage
\vspace{50cm}
\begin{figure}[b]
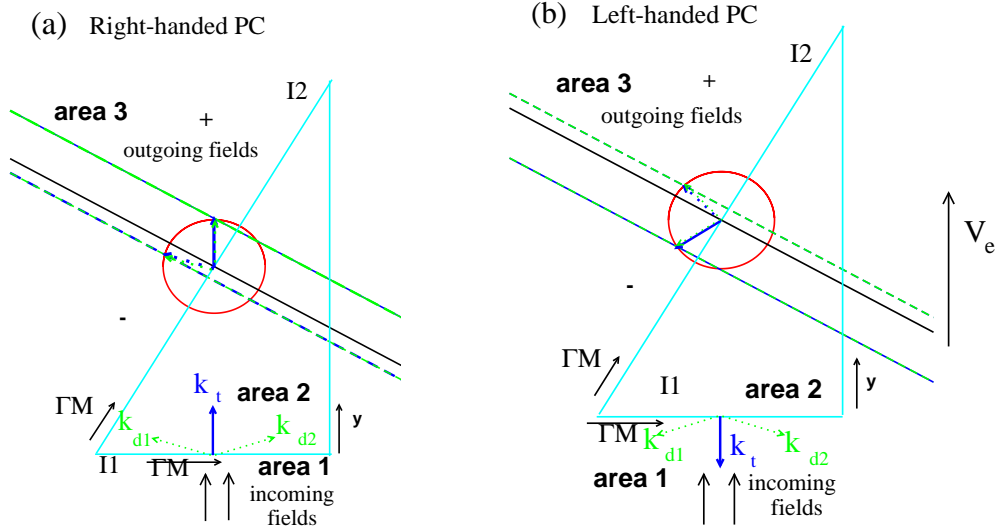
 
\begin{center}
\includegraphics[width=6.5cm,angle=0]{fig11a.eps} 
\includegraphics[width=6.5cm,angle=0]{fig11b.eps} 
\end{center}
\caption{ The wave vectors inside the PC wedge (area 2), and the outgoing wave vectors (area 3) for a case of anisotropic EFS. In (a) the wave vectors in area 2 are drawn assuming the PC is right-handed and in (b) assuming the PC is left-handed. Three different beams couple into the PC with three different wave vectors (${\bf k_{t}}, {\bf k_{d1}}, {\bf k_{d2}}$). The red circle represents the EFS in air for the relevant frequency ($\tilde f=0.50$). The bold and dotted blue and green lines perpendicular to I2 represent the different $k_{\|}$ values we obtain from the careful study of the wave vector diagram in the repeated zone scheme. We have four different outgoing beams. The position of the two is very close to the position of the remaining two. From the figure it becomes clear that the location of the beam with the larger angle coincides with the sign of ${\bf v_{e} \cdot k}$ inside the PC wedge and so determines the ``rightness.''}
\end{figure}

\begin{figure}[b]  
\begin{center}
\includegraphics[width=10cm]{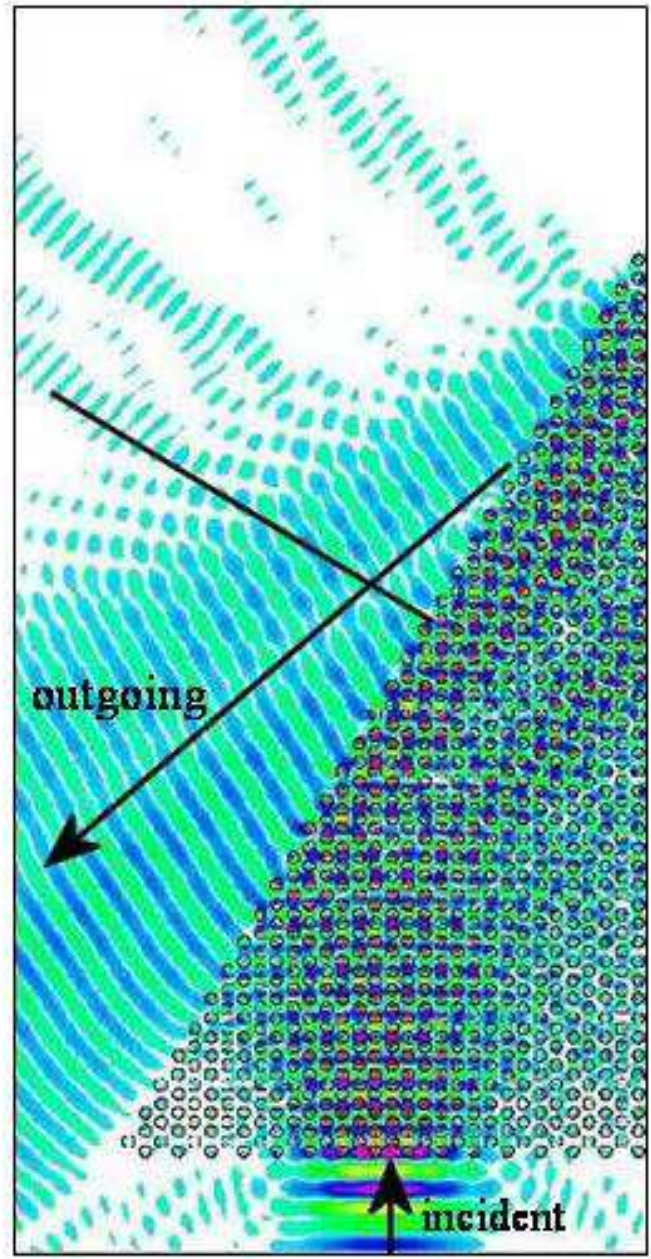}
\end{center}
\caption{Wedge simulation for the case of Fig 11. The outgoing beam with the larger angle is the negative hemisphere. Therefore, the system is left-handed in this case.} 
\end{figure} 
\begin{figure}[b]  
\begin{center}
\includegraphics[width=7.5cm]{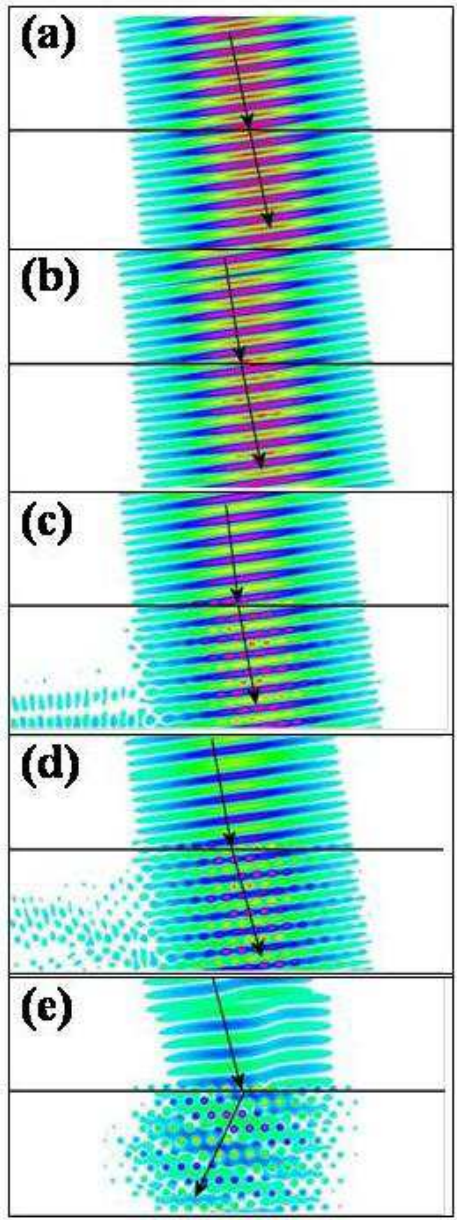}
\end{center}
\caption{Refraction at oblique incidence with angle of $8^{0}$ at a PC lattice of rods with radius 0.35 a for H-polarization. We consider a dielectric constant equal to 1.05, 1.2, 1.5, 2.0 and 5.0 for cases (a) through (e) respectively. Positive refraction is seen in all cases except in case (e) where index contrast is high.} 
\end{figure}
\begin{figure}[b] 
\begin{center} 
\includegraphics[width=12cm,angle=0]{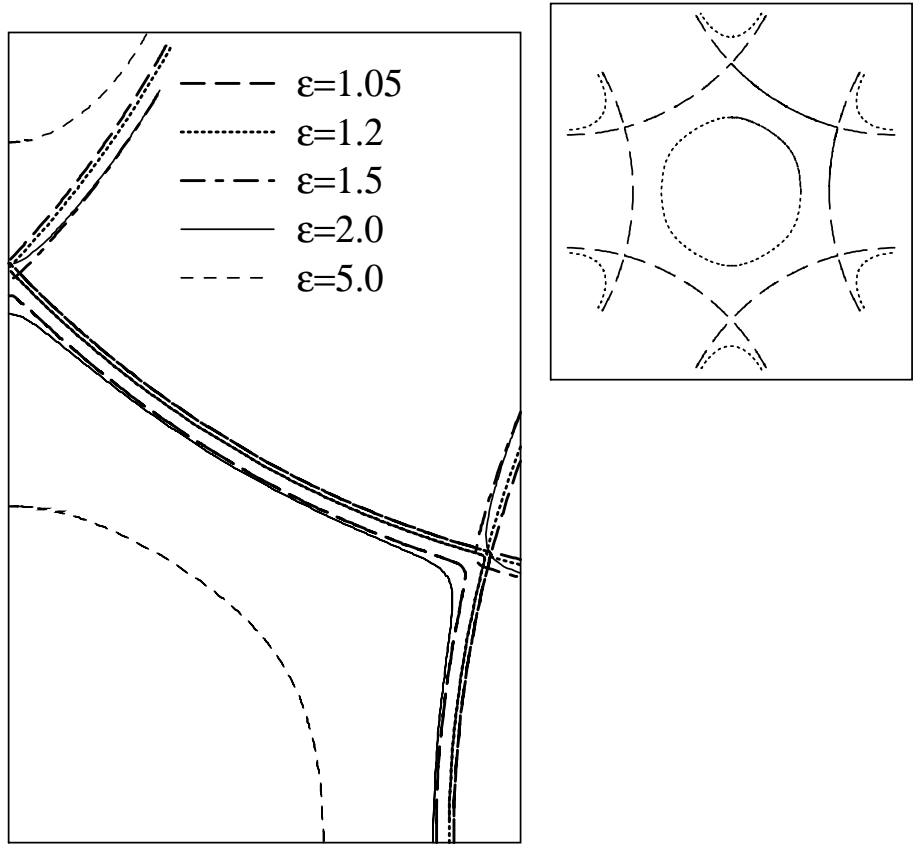} 
\end{center}
\caption{Equifrequency surfaces for 5 values of index contrast (1.05, 1.2, 1.5, 2.0 and 5.0 to 1.0 respectively). The operation frequency is chosen around the middle of the second and third band. The closed curves are the EFS that correspond to the second band, while the open concave-like curves are those that correspond to the third band. The EFS are drawn in the first quadrant only, so that we are able to see more detail. The third band EFS remain anisotropic with increasing index contrast. The ones that correspond to the second band become increasingly isotropic and finally almost circular for an index contrast as high as 5.0. In the right panel the EFS are drawn in the $2 \pi$-space for the limiting cases with $\epsilon$= 1.05 (dotted lines) and $\epsilon$=5.0. (dashed lines).}
\end{figure}
\begin{figure}[b] 
\begin{center} 
\includegraphics[width=12cm,angle=0]{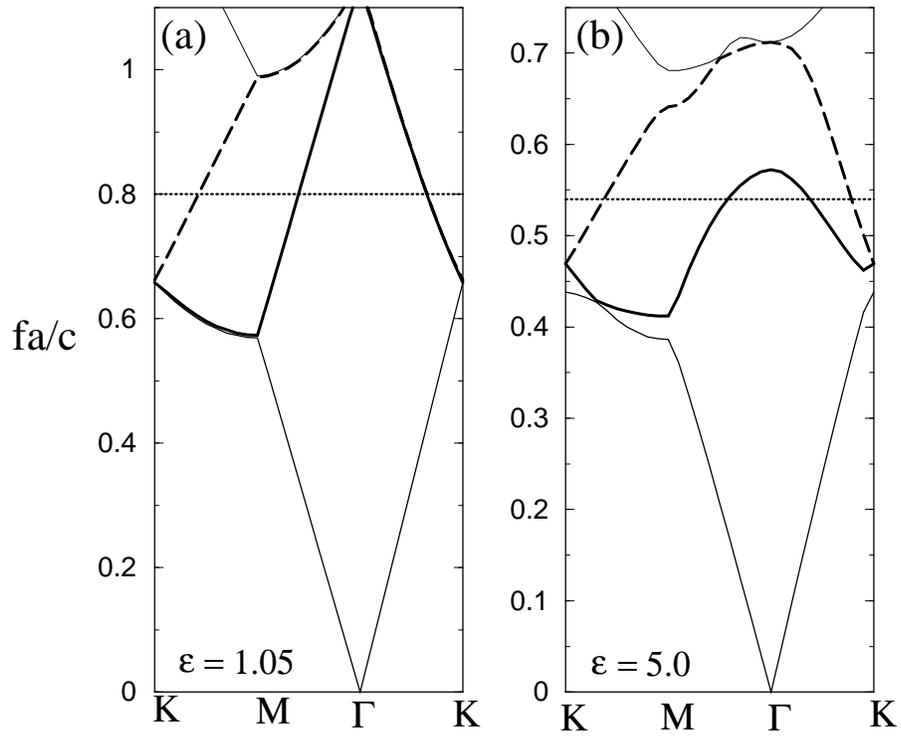} 
\end{center}
\caption{ The band structure for the two limiting cases with $\epsilon$= 1.05 and $\epsilon$=5.0. The operation frequency is chosen to be around approximately the middle of the second and third band, and is indicated in the figure with the dotted line.} 
\end{figure} 
\begin{figure}[b]  
\begin{center}
\includegraphics[width=13cm]{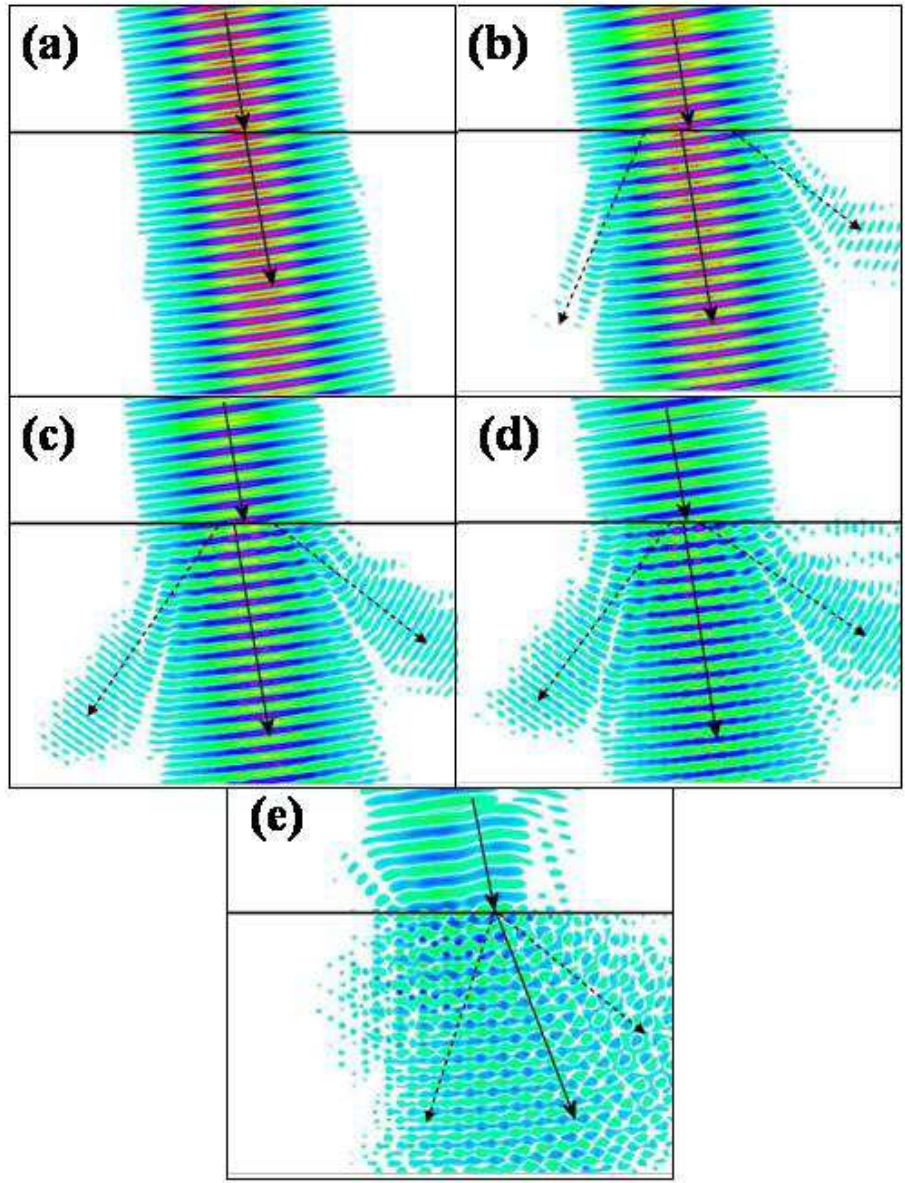} 
\end{center} 
\caption{Refraction at oblique incidence with an  angle of $8^{0}$ for the same cases as in Fig. 13, but with $\Gamma M$ as the symmetry direction of the interface. Even for an index contrast as low as 1.2, one can see three distinct beams (although two of them are faint in magnitude).}   
\end{figure}
\begin{figure}[b] 
\begin{center}  
\includegraphics[width=14cm,angle=0]{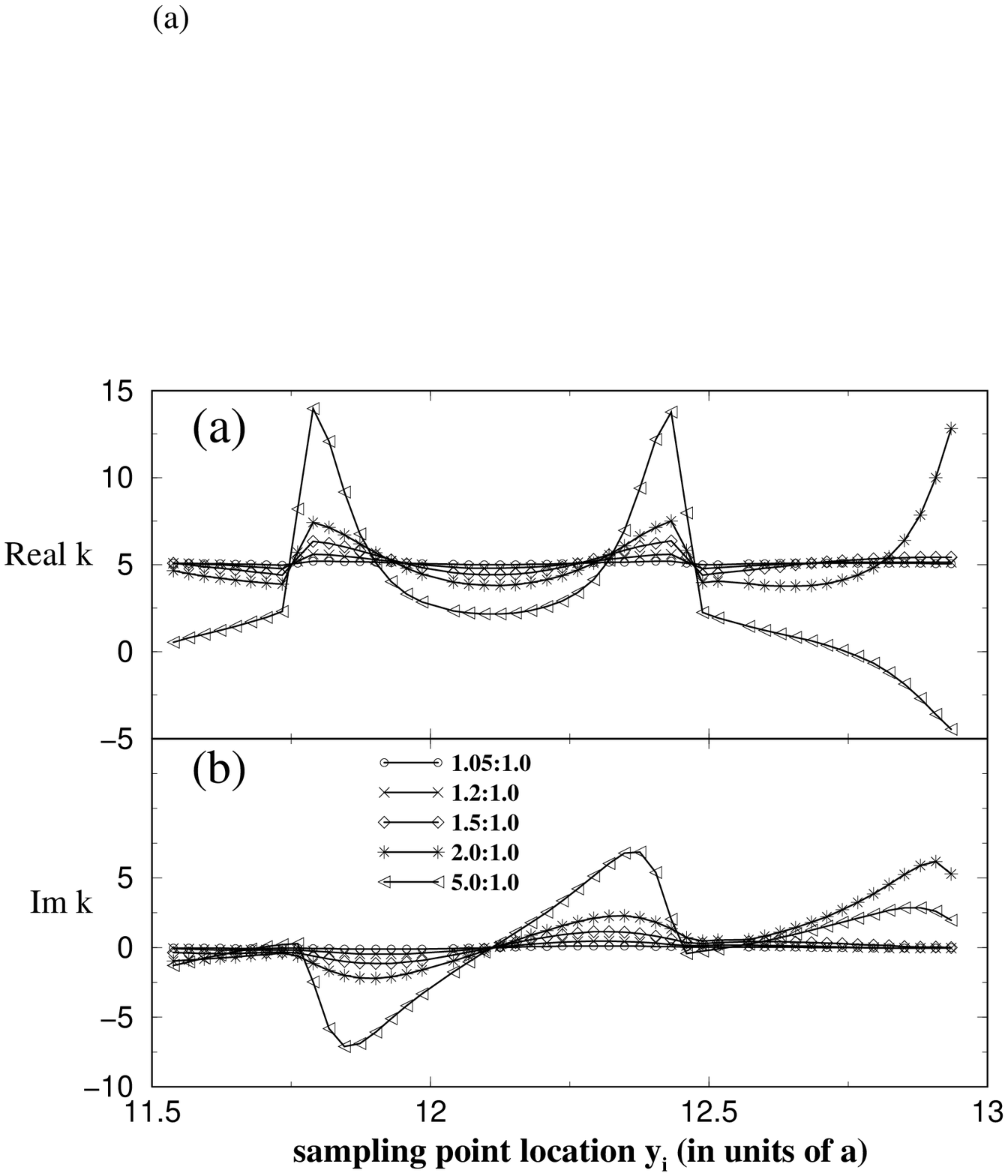} 
\end{center}
\caption{(a) Real  and (b) imaginary  part of the wave vector for the PC lattices of Figs. 13-16. The wave vector is calculated from the numerical FDTD field patterns at adjacent points $y_{i}$ and $ y_{i}+\Delta y$  ($\Delta y=\sqrt 3/62$ a). We have taken normal incidence along $\Gamma M$ and assumed in the wave vector extraction that one plane wave component dominates the propagation.}

\end{figure}


\begin{thebibliography}{1001}
\bibitem{note}
The PC structure is periodic in space. Therefore whenever we refer to $\vec S$ in the PC, the spatial average within the unit cell of the time-averaged Poynting vector is considered.
\bibitem{fundamental}
Usually, in the 1D literature the predominant wave vector of the Floquet-Bloch (FB) sum, is called as fundamental wave vector. In order to keep our analysis general, even for cases with very strong mixing in the FB sum, we choose the wave vector in the first zone to call as fundamental wave vector. See also Appendix I(b).
\bibitem{discrepancy}
The discrepancy is mainly due to the dicsretization in the FDTD simulation. In fact, if in the EFS analysis we use the discretized dielectric function instead, we obtain a wave vector value of $k=-2.26 \hspace{1mm} a^{-1}$.
\bibitem{Epol}
If the fields where E-polarized the quantity $A_{\eta}$ in Eq. (45) would be $\sum E_{\bf G}^2$. Thus, in this case too, it will be positive definite.
\end{thebibliography}
\end{document}